\documentclass[preprint,5p,twocolumn,11pt,sort&compress]{elsarticle}
\usepackage{graphicx}
\usepackage{bm}
\usepackage{amsmath}
\usepackage{amssymb}
\usepackage{amsfonts}
\usepackage{subfig}
\usepackage{hyperref}
\usepackage{bbold}
\hypersetup{colorlinks=true,citecolor=blue,linkcolor=blue,urlcolor=blue}
\usepackage[tableposition=top]{caption}
\usepackage{tabularx}
\usepackage{xspace}
\usepackage{multirow}
\usepackage[normalem]{ulem}
\usepackage{soul}
\usepackage[dvipsnames]{xcolor}
\usepackage[USenglish,british,american,australian,english]{babel}
\usepackage{comment}

\defcitealias{Euclid:2019clj}{EC19}

\usepackage{breakurl}
\usepackage{bbold}

\makeatletter

\usepackage{placeins}

\let\Gamma\varGamma
\let\Delta\varDelta
\let\Theta\varTheta
\let\Xi\varXi
\let\Pi\varPi
\let\Sigma\varSigma
\let\Upsilon\varUpsilon
\let\Phi\varPhi
\let\Psi\varPsi
\let\Omega\varOmega

\newcommand{\lcdm}{{\ensuremath{\Lambda\mathrm{CDM}}}\xspace}

\newcommand{\ee}{\end{equation}}
\newcommand{\ba}{\begin{eqnarray}}
\def\be{\begin{equation}}
\def\ee{\end{equation}}
\def\bea{\begin{eqnarray}}
\def\eea{\end{eqnarray}}

\def\de{\mathrm{d}}

\newcommand{\Q}

\DeclareRobustCommand{\ion}[2]{%
\relax\ifmmode
\ifx\testbx\f@series
{\mathbf{#1\,\mathsc{#2}}}\else
{\mathrm{#1\,\mathsc{#2}}}\fi
\else\textup{#1\,{\mdseries\textsc{#2}}}%
\fi}
\newcommand{\hi}{\ensuremath{\textrm{H\textsc{i}}}}

\definecolor{green}{rgb}{0, 0.69, 0.04}
\definecolor{red}{rgb}{1,0,0} 
\definecolor{magenta}{cmyk}{0,1,0,0} 
\definecolor{violet}{cmyk}{0,1,0,0} 
\definecolor{darkgreen}{rgb}{0,0.65,0.05}
\definecolor{antiquefuchsia}{rgb}{0.57, 0.36, 0.61}
\definecolor{burntorange}{rgb}{0.8, 0.33, 0.0}

\providecommand{\tabularnewline}{\\}

\newcommand{\planck}{\textit{Planck}}

\def\l@section{\@dottedtocline{1}{1em}{2em}}
\def\l@subsection{\@dottedtocline{2}{2em}{4em}}
\def\l@subsubsection{\@dottedtocline{3}{3em}{5em}}

\usepackage[titletoc]{appendix}

\newcommand{\Tstrut}{\rule{0pt}{2.6ex}}       

\usepackage{tikz}
\usepackage{collcell}

\newcommand\ColCell[1]{%
	\pgfmathparse{#1>0?1:0}%
	\ifnum\pgfmathresult=1\relax\color{red}\fi#1}

\newcolumntype{E}{>{\collectcell\ColCell}c<{\endcollectcell}}

\begin{document}

\begin{frontmatter}

\title{Constraining gravity with synergies between radio and optical cosmological surveys}

\author[Paris,Aachen]{Santiago Casas}
\author[Turin,INFNto]{Isabella P. Carucci}
\author[Paris]{Valeria Pettorino}
\author[Turin,INFNto,OATo]{Stefano Camera}
\author[OAR,INFN]{Matteo Martinelli}

\address[Paris]{Universit\'e Paris-Saclay, Universit\'e Paris Cit\'e, CEA, CNRS, AIM, 91191, Gif-sur-Yvette, France}
\address[Aachen]{Institute for Theoretical Particle Physics and Cosmology (TTK), RWTH
Aachen University, 52056 Aachen, Germany}
\address[Turin]{Dipartimento di Fisica, Universit\`a degli Studi di Torino, via P.\ Giuria 1, 10125 Torino, Italy}
\address[INFNto]{INFN -- Istituto Nazionale di Fisica Nucleare, Sezione di Torino, via P.\ Giuria 1, 10125 Torino, Italy}
\address[OATo]{INAF -- Istituto Nazionale di Astrofisica, Osservatorio Astrofisico di Torino, strada Osservatorio 20, 10025 Pino Torinese, Italy}
\address[OAR]{INAF -- Istituto Nazionale di Astrofisica, Osservatorio Astronomico di Roma, via Frascati 33, 00040 Monteporzio Catone, Italy}
\address[INFN]{INFN -- Istituto Nazionale di Fisica Nucleare, Sezione di Roma, piazzale Aldo Moro 2, 00185 Roma, Italy}

\date{\today}

\begin{abstract}
In this work we present updated forecasts on parameterised modifications of gravity that can capture deviations of the behaviour of cosmological density perturbations beyond \lcdm.
For these forecasts we adopt the SKA Observatory (SKAO) as a benchmark for future cosmological surveys at radio frequencies, combining a continuum survey for weak lensing and angular galaxy clustering with an \hi\ galaxy survey for spectroscopic galaxy clustering that can detect baryon acoustic oscillations and redshift space distortions.
Moreover, we also add 21cm \hi\ intensity mapping, which provides invaluable information at higher redshifts, and can complement tomographic resolution, thus allowing us to probe redshift-dependent deviations of modified gravity models. For some of these cases, we combine the probes with other optical surveys, such as the Dark Energy Spectroscopic Instrument (DESI) and the Vera C.\ Rubin Observatory (VRO). We show that such synergies are powerful tools to remove systematic effects and degeneracies in the non-linear and small-scale modelling of the observables. Overall, we find that the combination of all SKAO radio probes will have the ability to constrain the present value of the functions parameterising deviations from $\Lambda$CDM ($\mu$ and $\Sigma$) with a precision of $2.7\%$ and $1.8\%$ respectively, competitive with the constraints expected from optical surveys and with constraints we have on gravitational interactions in the standard model. Exploring the radio-optical synergies, we find that the combination of VRO with SKAO can yield extremely tight constraints on $\mu$ and $\Sigma$ ($0.9\%$ and $0.7\%$ respectively), which are further improved when the cross-correlation between intensity mapping and DESI galaxies is included.
\end{abstract}


\end{frontmatter}


\section{Introduction}
The current concordance cosmological model, \lcdm, has been able to pass a variety of tests along the years, and as of today it still is a very good fit to present available data (see \cite{Aghanim:2018eyx, 2016A&A...594A..14P, 2021arXiv210513549D, 2020A&A...633A..69H, 2019MNRAS.490.2155S, 2018MNRAS.476.4662V, 2018MNRAS.479.3422A} among many others). Despite the success of \lcdm, the nature of dark energy and dark matter remains unknown: from a theory point of view, there is no convincing prediction of the value of the cosmological constant $\Lambda$; it requires a high level of fine tuning in the initial conditions, and it marks our epoch as a very special time in the evolution of the Universe. Recent observations have highlighted tensions between low redshift measurements of cosmological parameters and their value inferred from high redshift observations through the assumption of a \lcdm expansion history (see e.g.\ \cite{Perivolaropoulos:2021jda} for a recent review). While such tensions may have an origin in unknown systematic effects, several works have been in parallel investigating whether scenarios alternative to \lcdm are able to overcome these shortcomings, also extending General Relativity (GR) beyond Einstein's theory, at cosmological scales (see \cite{2020arXiv200811284D} and references therein for a recent review.)

In this paper we rely on phenomenological parameterisations of departures from GR, and forecast our ability to test them with cosmological surveys with the upcoming SKA Observatory\footnote{\href{https://www.skao.int}{https://www.skao.int}} (SKAO), alone and in synergy with other surveys at optical/near-infrared wavelengths. We use parameterisations of the evolution of cosmological perturbations that modify the standard equations for the gauge-invariant gravitational potentials, $\Phi$ and $\Psi$. Perturbations are fully defined with two free functions of redshift and scale that modify the Poisson equation and the behaviour of the two gravitational potentials \cite{Kunz:2012aw,Amendola:2012ky}. While there is no unique choice for such free functions, we follow here the approach of \cite{planck_collaboration_planck_2016}, where the parameterised functions are $\mu$, modifying the Poisson equation for the Newtonian potential $\Psi$, and $\eta$, which determines the ratio between $\Phi$ and $\Psi$.

Even within this theoretical framework, there is no unique choice for such functions, and different approaches can be taken to parameterise them. For instance, one can assume they are scale-independent \cite{Aghanim:2018eyx}, or include extra parameters  controlling how these functions change with scale \cite{planck_collaboration_planck_2016,bull_extending_2015}, or use as free parameters the values of these functions in redshift bins \cite{Casas:2017eob,Pogosian:2021mcs,Raveri:2021dbu}. 
Extensions of this binned approach for parameterised modified gravity have been worked out in \cite{PhysRevD.101.123517} and implemented into $N$-body simulations in \cite{Srinivasan_2021}.
Moreover, a purely phenomenological investigation is not the only possible choice, and several results were obtained within the framework of the so-called effective field theory of dark energy \cite{Gubitosi2013}, which allows us to study departures from GR in the context of the Horndeski class of theories \cite{Gleyzes2016,Bellini2016,Alonso2016,Frusciante:2019xia}.

The common line of all these approaches is to study how departures from GR modify the evolution of cosmological perturbations. These studies will therefore particularly benefit of the increased sensitivity of galaxy surveys planned for the current decade (see e.g.\ \cite{Martinelli:2021hir} and references therein). Galaxy clustering (GC) and weak lensing cosmic shear (WL) data are particularly sensitive to modifications of the theory of gravity. The former probes the growth of cosmological structures and is sensitive to the evolution of $\Psi$, while the latter can probe the distribution of matter through its gravitational effects on the path of photons, and it is therefore sensitive to the combination $\Phi+\Psi$, which sources the lensing potential.

Furthermore, a new technique to probe cosmological structures has been advocated over the last decade: line intensity mapping (IM) \cite{Bharadwaj:2000av,Battye:2004re,Wyithe:2007rq,Chang:2007xk,2013PhRvL.111q1302C}. Doing IM of a particular galactic emission line means measuring the integrated radiation from unresolved sources in large patches of the sky. This way, we map the underlying dark matter field with excellent redshift resolution, making IM a sensitive probe of $\Psi$, and therefore a useful tool to constrain cosmological parameters and deviations from GR \cite{Carucci:2017cnn,Berti:2021ccw,Berti:2022ilk}. For instance, we can focus on the 21-cm line emitted by atomic neutral hydrogen (\hi), the most abundant baryonic element in the Universe and an optimal tracer of its structure. For cosmology, we are interested in the largest scales we can probe. Hence, we can perform \hi\ IM surveys with radio telescopes in the so-called single-dish mode. Each antenna/dish operates as a single telescope, not in interferometry, and maps are co-added. As a result, the angular resolution is low, but the area coverage unprecedented \cite{Chang2010,battyeps,Bull2015}.

In this work, we focus on the extensive radio surveys that the SKAO's Mid Telescope, located in the Karoo desert in South Africa, will be able to carry out. Thanks to these, we can exploit all the probes described above, through galactic radio continuum emission and 21-cm line emission from resolved galaxies and in IM \cite{SKA_cosmo_redbook}.

This paper is organized as follows: in \autoref{sec:parameterising-Modified-Gravity} we review the main equations used to describe phenomenologically deviations from GR and specify our choice of the parameterisation. In \autoref{sec:fisher} we present the Fisher matrix method used to obtain our forecasts, we describe the observational probes considered and highlight the experimental setup. Our forecast results are shown in \autoref{sec:results} and we draw our conclusions in \autoref{sec:conclusions}.

\section{\label{sec:parameterising-Modified-Gravity}Parameterising Modified Gravity}
We choose to work in the conformal Newtonian gauge and in a flat Universe, with the line element given by 
\begin{equation}
\de s^2=-(1+2\,\Psi)\,\de t^2+a^2\,(1-2\,\Phi)\,\de x_i\,\de x^i\;,
\end{equation}
where $a$ is the scale factor, related to the redshift $z$ via $1+z=1/a$. In this gauge, the two scalar metric perturbations $\Phi$ and $\Psi$, functions of time and scale, coincide with the gauge-invariant Bardeen potentials \cite{PhysRevD.22.1882}.

In theories with extra degrees of freedom (dark energy, DE) or modifications of General Relativity (modified gravity, MG), the normal linear perturbation equations are altered with respect to the standard case, thus leading to different values of $\Phi$ and $\Psi$ for a given matter source. Such departures from the standard behaviour of the two potentials can generally be encoded in two functions of time and scale. Several choices are possible and have been adopted in the literature for these functions, see e.g.\ \cite{planck_collaboration_planck_2016} for a limited overview. The choice we do in this work is to introduce the two functions through a modification of the Poisson equation for $\Psi$ and a gravitational slip. While the former changes the evolution in time and scale of the $\Psi$ potential, the latter introduces a difference between $\Psi$ and $\Phi$ (the equivalent of anisotropic stress) already at the linear level and for pure cold dark matter: \lcdm\  is retrieved when $\Psi=\Phi$.

The expressions that define $\mu(a,k)$ and $\eta(a,k)$ as the functions encoding the modified behaviour of the potentials are
\begin{align}
-k^2\,\Psi(a,k) & =4\,\pi\,G\,a^2\,\mu(a,k)\,\rho(a)\,\Delta(a,k)\;;\label{eq: mu_def}\\
\eta(a,k) & =\frac{\Phi(a,k)}{\Psi(a,k)}\;.\label{eq: eta_def}
\end{align}
Here $\rho(a)$ is the average dark matter density and $\Delta(a,k)\equiv\delta(a,k)+3\,a\,H(a)\,\bm\nabla\cdot\bm v(a,k)$ is the comoving density contrast with $\delta$ the fractional overdensity, $H$ the Hubble rate, and $\bm v$ the peculiar velocity
field. We neglect here relativistic particles and radiation as we are only interested in modeling the perturbation behaviour at late times. Under these assumptions, $\eta$, which is effectively a model independent observable \cite{amendola_observables_2013}, is closely related to modifications of GR via the gravitational potentials \cite{saltas_anisotropic_2014,sawicki_non-standard_2016}, while $\mu$ encodes deviations in gravitational clustering, especially in redshift-space distortions, as non-relativistic particles are accelerated by the gradient of $\Psi$.

In this work, we will also consider weak lensing observations, which are instead sensitive to deviations in the lensing or Weyl potential $\Upsilon=(\Phi+\Psi)/2$, since it is this combination that affects null-geodesics (relativistic particles). To this end we introduce a function $\Sigma(a,k)$ so that
\begin{equation}
-k^2\,\Upsilon(a,k)=4\,\pi\,G\,a^2\,\Sigma(a,k)\,\rho(a)\,\Delta(a,k)\;.\label{eq:Sigma-def}
\end{equation}
Note that, as such, $\Sigma$ plays the role of $\mu$ in a Poisson-like equation for the Weyl potential (cf.\ \autoref{eq: mu_def}). As metric perturbations are fully specified by two functions of
time and scale, this latter function $\Sigma$ is not independent from $\mu$ and $\eta$, and one can relate the three functions through
\begin{equation}
\Sigma(a,k)=\frac{\mu(a,k)}{2}\,\left[1+\eta(a,k)\right]\;.\label{eq:SigmaofMuEta}
\end{equation}
Throughout this work, we will denote the standard \lcdm model, defined through the Einstein-Hilbert action with a cosmological constant, simply as GR. For this case we have that $\mu=\eta=\Sigma=1$. All other cases in which these functions are not unity will be considered as MG models.

The advantage of using phenomenological functions such as  $\mu$ and $\eta$ is that they allow to model {\em any} deviations of the perturbation behaviour from \lcdm expectations, they are relatively close to observations, and they can also be related to other commonly used parameterisations \cite{2010PhRvD..81j4023P}.
On the other hand, they are not easy to map to an action (as opposed to approaches like effective field theories that are based on an explicit action) and in addition they contain so much freedom that we normally restrict their parameterisation to a subset of possible functions.

In this work we assume a simple parameterisation, based on the one used in the \planck\ analysis \cite{planck_collaboration_planck_2016}:
\begin{align}
    \mu(a,k)&=1+E_{11}\,\Omega_{{\rm
		DE}}(a)\;,\label{eq:DE-mu-parametrization}\\
    \eta(a,k)&=1+E_{22}\,\Omega_{{\rm
		DE}}(a)\;.\label{eq:DE-eta-parametrization}
\end{align}
This parameterisation is usually referred to as `late-time parameterisation', as it depends on the DE energy density $\Omega_{\rm DE}(a)$, and therefore allows a departure from GR mainly at low redshift where DE dominates.
We neglect here any scale dependence; the amplitude of the deviations from the GR limit is modulated by the parameters $E_{11}$ and $E_{22}$, while the time evolution of the MG functions is related to the DE density fraction.

For the forecasts presented below, we will show the constraints on $\mu$ and $\Sigma$ defined as the values that the functions defined in \autoref{eq:SigmaofMuEta} and \autoref{eq:DE-mu-parametrization} take at $z=0$, which in our parametrization is directly related to $\Omega_{{\rm DE,0}} \equiv \Omega_{{\rm DE}}(a=1)$.

\section{Fisher forecasts}\label{sec:fisher}
In this work we aim at forecasting the constraints that SKAO will be able to obtain on modifications of gravity. To achieve this goal we rely on a Fisher matrix analysis, and in this section we review its fundamentals, as well as how it can be applied to the observables of interest for the SKAO.

\subsection{Fisher formalism}\label{sec:fisherformalism}
Given a theoretical model describing a target observable and a set of experimental specifications for its measurement, the Fisher formalism provides us with a simple recipe to forecast marginal errors on the estimation of the model parameters. Starting from a likelihood function $L(\bm{\Theta})\equiv P(\bm d|\bm\Theta)$, representing the probability of the data, $\bm d=\{d_a\}$, given the model parameters $\bm{\Theta}=\{\Theta_{\alpha}\}$, the Fisher matrix \citep{Tegmark:1996bz, Euclid:2019clj} can be defined as
\begin{equation}
F_{\alpha \beta} =-\frac{\partial^2 \ln L(\bm{\Theta})}{\partial \Theta_{\alpha}\,\partial \Theta_{\beta}}\bigg |_{\rm fid} \;, \label{eq:Fish_def}
\end{equation}
where `fid' means that the derivatives are computed at the fiducial values of the model parameters, $\bm{\Theta}_{\rm fid}$.

Now, let us assume that $L(\bm{\Theta})$ is a multivariate Gaussian distribution, namely
\begin{multline}
-2\,\ln L(\bm\Theta) =\left[\bm d-\bm t(\bm\Theta)\right]^{\sf T}\,{\sf C}^{-1}\,\left[\bm d-\bm t(\bm\Theta)\right] \\
+\ln\det\left(2\,\pi\,{\sf C}\right)\;,\label{eq:Gaussian_like}
\end{multline}
where $\bm t(\bm\Theta)$ is the theoretical prediction, depending upon the model parameters, and ${\sf C}=\{C_{ab}\}$ is the data covariance matrix, which we assume does not depend on $\bm\Theta$.
Under these assumptions, \autoref{eq:Fish_def} applied to \autoref{eq:Gaussian_like} gives
\begin{equation}
F_{\alpha \beta} =\frac{\partial\bm t^{\sf T}}{\partial\Theta_\alpha}\,{\sf C}^{-1}\,\frac{\partial\bm t}{\partial\Theta_\beta}\; .\label{eq:Fish_general}
\end{equation}
In other words, the Fisher matrix is the inverse of the covariance matrix of the parameters. For this reason, it provides us with the expected errors around their fiducial values---in turn, an estimate of the ability of an experiment (or a combination of experiments) to constrain the parameters of the model.

In this work, we obtain our Fisher matrices using the \texttt{CosmicFish} code \cite{Raveri:2016xof,Raveri:2016leq}. We use an upgraded \texttt{python} implementation of this code that is not publicly available yet, but that will be released in the near future\footnote{The public version of \texttt{CosmicFish} is available at \url{https://cosmicfish.github.io/}.}. The cosmological functions used within \texttt{CosmicFish} to compute the observables are instead obtained from \texttt{MGCAMB} \cite{Zhao:2008bn,Hojjati:2011ix,Zucca:2019xhg}, which is able to obtain such functions in the MG model we consider in this work\footnote{In this work we use our own public fork of the \texttt{MGCAMB} repository, available at \url{https://github.com/santiagocasas/MGCAMB}.}.

\subsection{Spectroscopic galaxy clustering}\label{sec:GC}
GC probes the correlation among the three-dimensional positions of galaxies, which represent biased tracers of the distribution of matter in the Universe. The correlator of the Fourier transform of the matter density contrast at a given redshift $z$, $\delta_{\rm m}(z,\bm k)$, with itself is the matter power spectrum $P_{\delta\delta}(z,k)$. What we can measure through galaxy surveys, however, is the power spectrum of galaxies, rather than directly the one of matter. On large enough scales and in configuration space, the galaxy (number) density contrast $\delta_{\rm g}$ is related to that of matter through $\delta_{\rm g}=b_{\rm g}\,\delta_{\rm m}$, where $b_{\rm g}(z)$ is the so-called linear galaxy bias, and is assumed to be scale-independent in that regime. 

The cosmological information in GC is mostly contained in the shape of the baryon acoustic oscillations (BAO), which appear as wiggles in the power spectrum, and in the redshift space distortions (RSD), which induce anisotropies in galaxy number density fluctuations as a function of the angle with respect to the line of sight. While BAO are very sensitive to the baryonic content and the geometry of the Universe, RSD are very sensitive to the growth of density perturbations and the peculiar velocity field of matter and galaxies. In redshift space, we then write $\delta_{\rm g}=b_{\rm g}\,\delta_{\rm m}+(1+z)/H(z)\,\hat{\bm n}\cdot\bm\nabla(\hat{\bm n}\cdot {\bm v})$, with $\hat{\bm n}$ the line-of-sight direction and $\bm v$ the peculiar velocity field, whose radial component contributes to the measured redshift.

The observed power spectrum of galaxies is then given in terms of the matter power spectrum as \cite{Seo:2003pu,Seo:2007ns,Wang:2012bx}
\begin{multline} 
P_{\rm gg}(z,k,\mu_\theta)=  {\rm AP}(z)\times P_{\delta \delta, {\rm zs}}(z,k,\mu_\theta)\\
\times \exp\left\{-k^2\,\mu_\theta^2\,\left[\sigma_{z}^2(z)\,c^2/H^2(z)\right]\right\} \\ 
+ P_{\rm shot}(z) \;,\label{eq:P_obs}
\end{multline}
where $\mu_\theta\equiv\hat{\bm n}\cdot\bm k/k$, i.e.\ it is the cosine of the angle $\theta$ between the wave vector $\bm k$ and $\hat{\bm n}$. The first term in \autoref{eq:P_obs} corresponds to the Alcock-Paczynksi effect \cite{1979Natur.281..358A}, viz.\
\begin{equation}
{\rm AP}(z) \equiv \frac{\left[d_{\rm A,ref}(z)\right]^2\,H(z)}{d_{\rm A}^2(z)\, H_{\rm ref}(z)} 
\;,\label{eq:AP_eff}
\end{equation}
where $d_{\rm A}(z)$ is the angular diameter distance, and the subscript `ref' means that the corresponding quantity is calculated at the reference fiducial cosmology. The exponential term in \autoref{eq:P_obs} is a line-of-sight damping due to redshift uncertainty, modelled by its error $\sigma_z(z)$. Then, the additive term $P_{\rm shot}(z)$ is the extra contribution to account for incorrect subtraction of shot noise, which is usually set to zero. Lastly, $P_{\delta \delta, {\rm zs}}$ is the redshift-space power spectrum,
\begin{multline}
P_{\delta \delta, {\rm zs}}(z,k,\mu_\theta)=  {\rm FoG}(z, k, \mu_\theta)\times K_{\rm rsd}^2(b_{\rm g};z,k,\mu_\theta) \\
\times \frac{P_{\rm dw}(z,k,\mu_\theta)}{\sigma^2_8(z)} \;,\label{eq:P_zs}
\end{multline}
where the first term is due to non-linear RSD. It is called `Finger-of-God' (FoG) effect and models the damping of power on small scales due to the incoherent peculiar motions of galaxies,
\begin{equation}
   {\rm FoG}(z, k, \mu_\theta) \equiv \frac{1}{1+ k^2 \,\mu_\theta^2\, \sigma_{\rm p}^2(z)}\;.
\end{equation}
In the above equation, the strength of the FoG effect is modulated by the pairwise velocity dispersion, which we model as 
\begin{align}
\sigma_{\rm p}^2(z) &= \frac{1}{6\pi^2}\int\de k\, P_{\delta\delta}(k,z)f^2(k,z)\,,\label{eq:sigmap}
\end{align}
where $f\equiv{\rm d}\ln\delta/{\rm d}\ln a$ is the growth rate of matter perturbations and we have taken into account the possibility of a scale-dependent growth induced in a general modified gravity parametrization.
We compute this term at each redshift bin and evaluate it at the fiducial cosmology, keeping it fixed in our analysis, which corresponds to the optimistic settings in \cite{Euclid:2019clj}.
The $K_{\rm rsd}(b_{\rm g},z,k,\mu_\theta)$ term represents the Kaiser term, which accounts for linear redshift space distortions and is given by
\begin{equation} \label{eq:Krsd}
  K_{\rm rsd}(b_{\rm g};z,k,\mu_\theta) \equiv b_{\rm g}(z)\,\sigma_8 (z)+f(z, k)\,\sigma_8 (z)\, \mu_\theta^2\;.
\end{equation}

\begin{figure}
	\centering
\includegraphics[width=0.98\columnwidth]{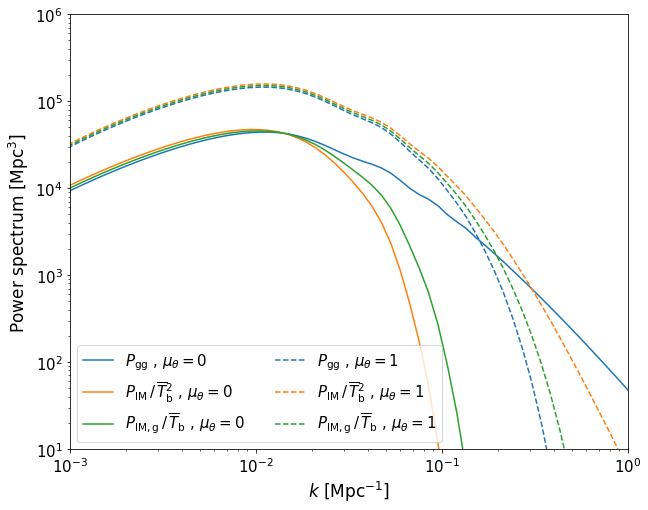}
	\caption{Galaxy power spectrum $P_{\rm gg}$ (blue lines), IM-\hi\ power spectrum $P_{\rm IM}$ (orange) and their cross-correlation $P_{\rm IM,g}$(green) as a function of scale for two angular directions $\mu=[0, 1]$ (solid and dashed, respectively), at $z=0.6$ (the lowest bin edge we consider for the combination of these probes). 
	The $P_{\rm IM}$ power spectrum has a strong damping at small scales in the perpendicular direction, since we include the effective beam in the signal, as shown in \autoref{eq:P_IM}.
    For the direction along the line of sight, the amplitude of the spectra is higher due to the Kaiser term, but the damping in $P_{\rm gg}$ is dominated by the FoG effect, as shown in \autoref{eq:P_obs}.
    }
	\label{fig:pggimnoise-ska1}
\end{figure}

Both the linear and non-linear RSD terms arise due to the transformation between redshift space and real space, when observing galaxies using redshift surveys. Here, $\sigma_8(z)$ the amplitude of matter fluctuations as a function of redshift. Finally, $P_{\rm dw}(z,k,\mu_\theta)$ stands for the `de-wiggled' power spectrum, modelling the effect of BAO damping on the matter power spectrum. We refer the reader to \cite{Euclid:2019clj} for a more detailed description of this term.
In \autoref{fig:pggimnoise-ska1} we plot the term $P_{\rm gg}(z=0.6,k,\mu_\theta)$ as a function of $k$ for two different values of $\mu_\theta$, namely $\mu_\theta=0$ and $1$ (solid and dashed blue  lines, respectively). This is a theoretical model of the galaxy power spectrum; dependence on a specific survey will enter in the spectroscopic redshift error $\sigma_z(z)$, which will be specified in \autoref{sec:method_and_data}.

A GC survey in a redshift bin of width $\Delta z$, centred on redshift $\bar z$ covers a volume $V_{\rm survey}$, and observes galaxies with comoving (volumetric) number density $N(z)$, depending on the survey specifications. The survey provides information for Fourier modes only in a range $\left[k_{\min}, k_{\max}\right]$, which also depends on the survey's specifications or on the scale until which one can accurately model non-linear scales.

Considering a GC survey carried out for a redshift range discretized into $N_{\rm b}$ redshift bins, we evaluate $P_{\rm gg}(z, k, \mu; \bm{\Theta})$ and its derivatives at the centre $\bar z_{m}$ of each redshift bin $m$, and at the fiducial value for each of the $N_\theta$ cosmological parameters. While we may expect the Fisher matrix to be an $ N_\theta \times N_\theta$ matrix, it is in practice more complicated due to the presence of $b_{\rm g}(z)$ and $P_{\rm shot}(z)$, which are in general unknown. In order to address this problem, we discretize $b_{\rm g}(z)$ and $P_{\rm shot}(z)$ into $N_{\rm b}$ redshift bins, assuming them to be mutually independent and considering them at each redshift bin as additional independent model parameters with some fiducial values. Thus our full Fisher matrix is of dimension $( N_\theta+2\,N_{\rm b} ) \times ( N_\theta+2\,N_{\rm b})$, and we can in the end marginalize it over the $2\,N_{\rm b}$ nuisance parameters.

Given the full (cosmological + nuisance) parameter set $\bm{\Theta}=\left\{\theta_{\alpha}, b_{{\rm g},m}, P_{{\rm shot}, m}\right\}$, where $\theta_\alpha$ are the cosmological parameters, $b_{{\rm g},m}\equiv b_{\rm g}(\bar z_{m})$ and $P_{{\rm shot}, m}\equiv P_{\rm shot}(\bar z_{m})$, the total Fisher matrix for a GC survey over all redshift bins can be written as \citep{Euclid:2019clj}
\begin{multline}
    F^{AB}_{\alpha\beta}=\sum_{m,n=1}^{N_{\rm b}}\sum_{a,b,c,d,n}\frac{\partial P_{AB}(\bar z_m,k_a,\mu_b)}{\partial \Theta_{\alpha}} 
    \\\times\frac{\partial P_{AB}(\bar z_n,k_c,\mu_d)}{\partial \Theta_{\beta}}\, \left[{\sf C}^{AB}(\bar z_m,\bar z_n)\right]^{-1}_{abcd}\;,\label{eq:GCFisher}
\end{multline}
where $A,B$ label the probe under scrutiny, i.e.\ $A=B={\rm g}$ for galaxy clustering. Above, $k_a$ and $\mu_b$ represent the discretised values of $k$ and $\mu_\theta$ the signal has been binned into, and ${\sf C}$ is the covariance matrix between a set of measurements of $P_{AB}(\bar z_m,k_a,\mu_b)$ and one of $P_{AB}(\bar z_n,k_c,\mu_d)$. Again, in full generality, it reads
\begin{multline}
    {\sf C}_{abcd}^{AB}(\bar z_m) =
    \frac{4\,\pi^2\,\delta^{\rm K}_{ac}\,\delta^{\rm K}_{bd}\,\delta^{\rm K}_{mn}}{k_a^2\,\Delta k_a\,\Delta\mu_b\,V_{\rm survey}}\\
    \times \Big[\tilde P_{AA}(\bar z_m,k_a,\mu_b)\,\tilde P_{BB}(\bar z_m,k_a,\mu_b)\\
    + \tilde P_{AB}(\bar z_m,k_a,\mu_b)\,\tilde P_{AB}(\bar z_m,k_a,\mu_b)\Big]\;,\label{eq:GC_Cov_IMg}
\end{multline}
where $\tilde P_{AB}=P_{AB}+P_{AB,{\rm noise}}\,\delta^{\rm K}_{AB}$.

The power spectrum and its derivatives appearing in \autoref{eq:GCFisher} are evaluated at the fiducial values of the parameters, and the final Fisher matrix is the combination of Fisher matrices at different redshift bins, i.e.\ the sum of the $N_{\rm b}$ Fisher matrices. We also marginalize over the irrelevant parameters at this stage to obtain a matrix of dimension $ N_\theta \times N_\theta$ as the resulting $F^{\rm GC}_{\alpha\beta}$ for the cosmological parameters, which contains the constraint information about the parameter set $\bm\theta$.

We want to stress here that in order to follow this approach one needs to have very precise measurements of the redshifts of the galaxies. Such a precision can be achieved using spectroscopic measurements and, therefore, we will refer to this observational probe as `GCsp' throughout the rest of the paper. This will avoid confusion with other probes of galaxy correlations (see \autoref{sec:WL}).

\subsection{Intensity Mapping}

\hi\ emits 21-cm radiation due to its spin-flip transition, which can be detected with the IM technique. To model the power spectrum of the IM signal, we need to take into account the large-scale distribution of \hi\ in the cosmic epochs we are considering. Following the literature \citep{Furlanetto:2006jb}, the observed 21-cm average brightness temperature at a given redshift is given by
\begin{equation}
\bar{T}_{\rm b}(z)= 189\,h\,\frac{(1+z)^2\,H_0}{H(z)}\,\Omega_{\rm HI}(z) \,\mathrm{mK}\;,
\label{eq:Tbar}
\end{equation}
which implies that, the larger the amount of \hi, the larger the amplitude of the signal. In turn, the latter is determined by the cosmic \hi\ comoving density fraction, $\Omega_{\rm HI}(z)$. After reionization ($z \lesssim 6$), neutral gas mostly resides in the densest regions of the cosmic web---dark matter haloes \citep[][]{bagla2009, Carucci:2015bra,Carucci:2017cnn, illustris_IM}. Thus, we can consider \hi\ a biased tracer of the underlying matter density field, and we express the total 21-cm brightness temperature at a given redshift and in a unit direction $\hat{\bm n}$ in the sky as \cite{SKA_cosmo_redbook}  
\begin{multline}\label{eq:Tb}
T_{\rm b}(z,\hat{\bm n}) = \bar{T}_{\rm b}(z)\,\bigg[1+b_{\rm HI}(z)\,\delta_{\rm m}(z,\hat{\bm n})\\
-\frac{(1+z)}{H(z)}\,\hat{\bm n}\cdot\bm\nabla\left(\hat{\bm n}\cdot {\bm v} \right)\bigg]\;,
\end{multline}
where $b_{\rm HI}$ is the \hi\ bias, $\delta_{\rm m}$ is the matter density contrast, and $\bm v$ is the peculiar velocity of the \hi\ clouds, giving rise to RSD.

Given the relations in \autoref{eq:Tbar} and \autoref{eq:Tb}, it follows that we can model the 21-cm signal once we have a prescription for $\Omega_{\rm HI}$ and $b_{\rm HI}$. Abundance and clustering properties of \hi\ have been measured by local-Universe \hi\ galaxies surveys \citep[e.g.][]{alfalfa} and through the \hi\ column densities of absorption systems present in the spectra of quasars \citep[e.g.][]{Crighton2015}. As suggested by \cite{Carucci2020}, we make use of the aforementioned compilations and define
\begin{align}
b_{\rm HI}(z) &= 0.3\,(1+z) + 0.6 \; , \\
\Omega_{\rm HI}(z)  &= 4.0\,(1+z)^{0.6} \times 10^{-4}\;.
\end{align}


Then, we define the power spectrum of the 21-cm  signal in IM as
\begin{equation}\label{eq:P_IM}
P_{\rm IM}(z,k) = \bar{T}_{\rm b}^2(z)\, {\rm AP}(z)  \,P_{\delta \delta, {\rm zs}}(z,k,\mu_\theta)\, \beta^2(z,k,\mu_\theta) \;,
\end{equation} 
where the first term is the average brightness temperature illustrated above with \autoref{eq:Tbar}, the second is the Alcock-Paczynksi term introduced in \autoref{eq:AP_eff}, the third is the redshift-space power spectrum as defined in \autoref{eq:P_zs} but replacing the galaxy bias with the \hi\ bias $b_{\rm HI}$ in the $K_{\rm rsd}$ term of \autoref{eq:Krsd}. Finally, the fourth term $\beta(z,k,\mu_\theta)$ is the effective telescope beam that dumps the power at scales below that of the resolution of the maps. We model the latter as \citep{Bull2015,Alonso2016}
\begin{equation}
    \beta(z,k,\mu_\theta) = \exp\left[-\frac{k^2\,(1-\mu_\theta^2)\,r^2(z)\,\theta_{\rm pb}^2(z)}{16\,\ln 2} \right]\;.
\end{equation}
Above, $r(z)$ is the comoving distance to redshift $z$ and $\theta_{\rm pb}$ is the full width at half maximum of the dish primary-beam, which we model as \citep{Matshawule2020}
\begin{equation}
    \theta_{\rm pb}(z) = 1.22\, \frac{\lambda(z)}{D_{\rm d}}\;,
\end{equation}
with $\lambda(z) = (1+z)\times21\,\mathrm{cm}$ the wavelength of the observed (redshifted) frequency, and $D_{\rm d}$ the diameter of the telescope dish.
In \autoref{fig:pggimnoise-ska1} we plot the term $P_{\rm IM}(z,k)$ divided by the brightness temperature $\bar{T}_{\rm b}^2(z)$ as a function of $k$ and at $z=0.6$ (the lowest $z$-bin considered for IM) for two different values of $\mu_\theta$, namely $\mu_\theta=0$ (solid orange line) and $\mu_\theta=1.0$ (dashed orange line). As mentioned above, the effective beam is part of the signal for $P_{\rm IM}$, which dampens it considerably at small scales in an angle-dependent way. For $\mu_\theta=0$, which corresponds to a $90$ deg angle with respect to the line of sight, the damping kicks in already at scales as large as $k \approx 0.03\,\mathrm{Mpc}^{-1}$, which means that most of the information we obtain comes from modes along the line of sight.  

We notice that the amplitude of the $P_{\rm IM}$ power spectrum depends through the $\bar{T}_{\rm b}(z)$ term on both the overall amount of neutral hydrogen in the Universe at a given redshift, $\Omega_{\rm HI}(z)$, and its relation to the underlying dark matter density field, through $b_{\rm HI}(z)$. 

Taking into account all different contributions, the cross-correlation power spectrum of IM with a galaxy sample with bias $b_{\rm g}$ and cross-correlation coefficient $r_{\rm HI,g}$ reads
\begin{align}\label{eq:PIMg-cross}
P_{\rm IM,g}(z,k, \mu_\theta) &=
r_{\rm HI,g} \, {\rm AP}(z)\, {\rm FoG}(z, k, \mu_\theta) \, \bar{T}_{\rm b}(z)\nonumber \\ 
&\quad\times K_{\rm rsd}(b_{\rm g}, z, k)\,K_{\rm rsd}(b_{\rm HI}, z, k) \nonumber\\
&\quad\times \frac{P_{\rm dw}(z,k,\mu_\theta)}{\sigma^2_8(z)}\, \beta^2(z,k,\mu_\theta) \;.
\end{align}
It has a subdominant shot-noise contribution that can be safely neglected \cite{Castorina2017,illustris_IM} and has intrinsically exquisite redshift resolution. 
We model the noise power spectrum of IM as \cite{Bull2015}
	\begin{equation}
	{P}_{\rm noise}(z) = \frac{2\pi \, f_{\rm sky}}{\nu_{21}(z) \, t_{\rm tot} \, N_{\rm d}}\,\frac{(1+z)^2 \, r^2(z) }{H(z) }\,\left[\frac{T_{\rm sys}(z)}{\bar{T}_{\rm b}(z)}\right]^2\;, 
	\label{eq:Pnoise}
	\end{equation}
where $f_{\rm sky}$ is the observed sky fraction, $\nu_{21}(z)$ is the emission frequency corresponding to the 21-cm line at redshift $z$, $t_{\rm tot}$ is the total observing time, $N_{\rm d}$ the number of dishes used for the observation, and $T_{\rm sys}$ is the system temperature. 

Having the IM power spectrum in \autoref{eq:P_IM} and its cross-correlation with galaxy tracers in \autoref{eq:PIMg-cross}, we can now use these in \autoref{eq:GCFisher} with the covariance matrix defined in \autoref{eq:GC_Cov_IMg} to forecast the constraining power of this probe.

In this work, we consider that the IM survey is performed in \textit{single-dish mode}, which better suits cosmological studies with respect to standard radio-interferometry in the SKAO set-up \citep[e.g.\ see discussion in][]{Bull2015}, and was successfully applied for the first time in \citep{Wang:2020lkn,Cunnington:2022uzo}. We assume that astrophysical foregrounds and systematics have been successfully removed from data: although the cleaning of these contaminants has been the bottleneck of IM surveys, it is now an active line of research and progress is ongoing (e.g.\ recent work by \cite{Carucci2020,Cunnington2021,SKAchallenge}).
For this reason, to date only detections in cross-correlation with galaxy surveys have been made \citep[see][]{2010Natur.466..463C, Anderson:2017ert, Wolz:2021ofa, Cunnington:2022uzo, CHIME:2022kvg}.

\subsection{Angular probes}\label{sec:WL}
The probes discussed up to now allow to use the full three-dimensional information encoded in galaxy surveys to reconstruct the matter power spectrum. However, this requires an extreme precision in the measurement of galaxy redshifts in order to be feasible, a precision that is not always available in observations of the large-scale structure. If that is the case, one can rely instead on what we refer here as `angular probes'. In this case, one compares the two-dimensional angular power spectrum of the observables, expanded in harmonic space, and binned in redshift. Such an approach is commonly used for correlations of galaxy shapes, i.e.\ for WL and GC. In order to distinguish the approach of this section from the one of \autoref{sec:GC}, we refer to the latter as GCph or GCco, dependending on the technique used to obtain the measurements, i.e.\ from photometric or radio-continuum observations, respectively. We use the same naming convention also for WL, labeling such a probe as WLco if obtained from continuum radio observations and WLph if coming from photometric measurements. We will discuss the specific surveys later on in \autoref{sec:method_and_data}.

A WL survey infers gravitational lensing of the light emitted from distant galaxies due to the distribution of matter along the line of sight by measuring correlations in the orientation of the galaxies. WL surveys probe, simultaneously, both the geometry of the Universe and the growth of structure through the matter power spectrum. By measuring the correlations in the image distortions of galaxies, one can reconstruct the matter density field. The target summary statistics is the angular power spectrum of the weak lensing effect of cosmic shear, $\gamma$, which reads
\begin{equation}\label{eq:cijdef_shear}
C_{ij}^{\gamma\gamma}(\ell) = 
\int\de z\,\frac{W_i^\gamma(z)\,W_j^\gamma(z)}{H(z)\,r^2(z)}\,P_{\Upsilon\Upsilon}(z,k_\ell)\;.
\end{equation}
The indices $i$ and $j$ denote the redshift bins of a tomographic WL survey, allowing us to use the information on the time evolution of $\Upsilon$ provided by the survey. Finally, the quantity $W_i^\gamma(z)$ is the so-called lensing kernel, a purely geometrical quantity given, in a flat Universe, by
\begin{equation}\label{eq:hatw}
W_i^\gamma(z) = r(z)\int_z^\infty\de z'\,\frac{r(z')-r(z)}{r(z')}\,n_i(z')\;,
\end{equation}
where $n_i(z)$ is the physical (surface) number density of galaxies in the $i$th redshift bin.\footnote{Note that the relation between the comoving volumetric number density and the physical surface number density is $n(z)=r^2/H\,N(z)$, given $\de V=r^2/H\,\de z\,\de\Omega$, with $\de\Omega$ the solid angle \citep[see e.g.][]{2021JCAP...12..009M}.} This can be obtained by convolving the redshift distribution of galaxies with the redshift measurement errors, which we model with the sum of two exponentials as in \cite{Kitching:2008eq}, with the parameter $\sigma_z^{\rm ph}$ determining the observational error on the redshift of the sources.

Notice that in order to express the shear power spectrum in the form of \autoref{eq:cijdef_shear}, the Limber and flat-sky 
approximations \citep{Kaiser:1991qi,LoVerde:2008re,Kitching:2016zkn,Kilbinger:2017lvu,Lemos:2017arq,Matthewson:2020rdt,2022MNRAS.510.1964M} have been used. They allow us to relate a wavenumber $k$ and a multipole $\ell$ through $k_\ell=(\ell+1/2)/r(z)$ \citep[see e.g.][ for the full and exact computation]{Taylor:2018qda}. 

The Weyl power spectrum $P_{\Upsilon\Upsilon}$ is related to the matter power spectrum $P_{\delta\delta}$ by
\begin{equation}\label{eq:weylLAM}
\frac{P_{\Upsilon\Upsilon}(z,k)}{P_{\delta\delta}(z,k)}=\Sigma^2(z,k)\, \left[\frac32\,H_0^2\,\Omega_{{\rm m},0}\,(1 + z)\right]^2\;.
\end{equation}
We can therefore use this relation to express \autoref{eq:cijdef_shear} in terms of $P_{\delta\delta}$ as
\begin{equation}
C_{ij}^{\gamma\gamma}(\ell) = 
\int\de z\,\frac{\hat W_i^\gamma(z)\,\hat W_j^\gamma(z)}{H(z)\,r^2(z)}\,P_{\delta\delta}(z,k_\ell)\;,
\end{equation}
where the new kernel function is given by
\begin{equation}
   \hat W_i^\gamma(z) = \frac94\,H_0^4\,\Omega_{\rm m,0}^2\,(1+z)^2\,\Sigma^2(z,k)\,W_i^\gamma(z)\;.
\end{equation}

We must stress at this point that the observed distortion of distant galaxy images is not produced only by the shear we modeled through \autoref{eq:cijdef_shear}. An additional contribution comes from the intrinsic alignment (IA) of galaxies---an effect that contributes to overall ellipticity power spectrum $C_{ij}^{\epsilon\epsilon}(\ell)$. We model this contribution following \cite{Euclid:2019clj}, and we can therefore compute the full distortion power spectrum, which can be compared with observations, as
\begin{equation}\label{eq:cijdef}
C_{ij}^{\epsilon\epsilon}(\ell) = 
\int\de z\,\frac{\hat W_i^\epsilon(z)\,\hat W_j^\epsilon(z)}{H(z)\,r^2(z)}\,P_{\delta\delta}( z,k_\ell)\;,
\end{equation}
with $\hat W_i^\epsilon(z)$ the combined shear and IA kernel, viz.\
\begin{equation}
    \hat W_i^\epsilon(z) =\hat W_i^\gamma(z)- \frac{\mathcal{A}_{\rm IA}\,\mathcal{C}_{\rm IA}\,\mathcal{F}_{\rm IA}(z)}{D(z)}\,n_i(z)\,H(z)\;.
    \label{eq:window-lensing}
\end{equation}
Above, $\mathcal{F}_{\rm IA}(z)=(1+z)^{\eta_{\rm IA}}L(z)^{\beta_{\rm IA}}$, with $\mathcal{A}_{\rm IA}$, $\beta_{\rm IA}$, and $\eta_{\rm IA}$ being IA nuisance parameters, $L(z)$ is the luminosity function of the observed galaxies, and $\mathcal{C}_{\rm IA}=0.0134$.

Similarly, we can obtain theoretical predictions for the observations of the galaxy position correlation function, i.e.\ what is observed by GCph and GCco surveys. The galaxy angular power spectrum can be obtained as 
\begin{equation}\label{eq:cijdef_gcph}
C_{ij}^{\rm gg}(\ell) = 
\int\de z\,\frac{\hat W_i^{\rm g}(z)\,\hat W_j^{\rm g}(z)}{H(z)\,r^2(z)}\,P_{\delta\delta}(z,k_\ell)\;.
\end{equation}
The galaxy clustering kernel is given by
\begin{equation}
    \hat W_i^{\rm g}(z) = b_i(z)\,n_i(z)\,H(z)\;,
    \label{eq:window-galaxy}
\end{equation}
with $b_i(z)$ the linear galaxy bias, which we model following the approach of \cite{Euclid:2019clj}, thus introducing a free parameter $b_i$ for each of the redshift bins.

Given that the observed galaxies used for WLph-GCph and WLco-GCco come from the same galaxy population, it is natural to expect that the cross-correlation (XCph and XCco) between these two observables, $C_{ij}^{\epsilon{\rm g}}$, will be non-vanishing. Throughout the rest of the paper we can therefore consider all angular probes together (also known as  3$\times$2pt analysis) thus using combinations GCco+WLco+XCco and GCph+WLph+XCph. The generic correlation can be theoretically modeled as
\begin{equation}\label{eq:cijdef_3x2pt}
C_{ij}^{ab}(\ell) = 
\int\de z\,\frac{\hat W_i^a(z)\,\hat W_j^b(z)}{H(z)\,r^2(z)}\,P_{\delta\delta}(z,k_\ell)\;,
\end{equation}
with $a,b=\{\epsilon,{\rm g}\}$.

The full Fisher matrix of the full combination of observables can be written, under the assumption of a Gaussian likelihood as in \autoref{eq:Gaussian_like} and that the data covariance matrix does not depend on the model parameters, as \cite{Euclid:2019clj}
\begin{multline}
    F_{\alpha\beta}^{AB}=\,\sum_{\ell,\ell'=\ell_{\rm min}}^{\ell_{\rm max}}\,\sum_{i,j,m,n}\,\frac{\partial C^{AB}_{ij}(\ell)}{\partial \Theta_{\alpha}}\,\frac{\partial C^{AB}_{mn}(\ell)}{\partial \Theta_{\beta}}\\
    \times\left[{\sf C}^{AB}(\ell,\ell')\right]^{-1}_{ijmn}\;,
\label{eq:3times2pt_Fisher}
\end{multline}
where indexes $A,B=\{\epsilon,{\rm g}\}$, while $i,j,m,n=1\ldots N_{\rm b}$. The covariance matrix between a measurements $C^{AB}_{ij}(\ell)$ and $C^{AB}_{mn}(\ell')$ is given by
\begin{multline}
    {\sf C}^{AB}_{ijmn}(\ell,\ell')=\frac{\delta_{\ell\ell'}^{\rm K}}{(2\ell+1)\,f_{\rm sky}\,\Delta\ell}\\
    \times\left[\tilde C_{im}^{AB}(\ell)\,\tilde C_{jn}^{AB}(\ell)+\tilde C_{in}^{AB}(\ell)\,\tilde C_{jm}^{AB}(\ell)\right]\;,
    \label{eq:3times2pt_Cov}
\end{multline}
with $\delta^{\rm K}$ the Kronecker delta, $\Delta\ell$ the width of the multipole bin(s), and $\tilde C^{AB}_{ij}(\ell)=C^{AB}_{ij}(\ell)+N^{AB}_{ij}$ (cf.\ \autoref{eq:GC_Cov_IMg}). The noise terms read
\begin{align}
N_{ij}^{\epsilon\epsilon} &= \frac{\epsilon_{\rm int}^2}{\bar{n}_i}\delta^{\rm K}_{ij}\;,\nonumber\\
N_{ij}^{\rm gg} &= \frac{1}{\bar{n}_i}\delta^{\rm K}_{ij}\;,\nonumber\\
N_{ij}^{\epsilon{\rm g}} &= 0\;,
\label{eq:photonoise}
\end{align}
where $\epsilon_{\rm int}$ is the intrinsic galaxy ellipticity scatter and $\bar{n}_i$ is the galaxy surface density in the $i$-th bin. \citep[for details, see][]{Casas:2015qpa,Casas:2017wjh,Amendola:2016saw}).

The high-multipole cutoff $\ell_{\rm max}$ in \autoref{eq:3times2pt_Fisher} encodes our ignorance of clustering, systematics and baryon physics on small scales; we discuss our choice for this in \autoref{nonlinear}.

\subsection{Fiducial cosmology, analysis and data}\label{sec:method_and_data}

This work aims to forecast the constraining power on possible departure from GR that the probes described in this section will bring, using the Fisher matrix formalism we summarised.

We do so by obtaining the forecast bounds on the $\mu$ and $\eta$ functions introduced in  \autoref{sec:parameterising-Modified-Gravity}, with the $E_{11}$ and $E_{22}$ parameters determining the amplitude of the deviation from GR in \autoref{eq:DE-mu-parametrization} and \autoref{eq:DE-eta-parametrization}. These are added to the set of free parameters of a standard cosmological analysis, i.e.\ the baryonic and total matter-energy densities $\Omega_{{\rm b},0}$ and $\Omega_{{\rm m},0}$, the reduced Hubble constant $h=H_0/100$, the tilt of the primordial power spectrum $n_{\rm s}$ and the root mean square of present-day linearly evolved density fluctuations in spheres of $8\,h^{-1}\,\mathrm{Mpc}$ radius $\sigma_8$. We indicate with $\theta_{\rm cosmo}$ the full set of free cosmological parameters, i.e.\ $\theta_{\rm cosmo}=\{\Omega_{{\rm m},0},\Omega_{{\rm b},0},h,n_{\rm s}, \sigma_8, E_{11},E_{22}\}$. 
We use as fiducial values for these the mean values obtained in \cite{planck_collaboration_planck_2016} (see \autoref{tab:cosmopars-fiducial}), which we will refer to as Planck15. The values for $E_{11}$ and $E_{22}$ depart significantly from the GR limit $E_{11}=E_{22}=0$ as CMB data alone are not able to tightly constrain them. Using this fiducial cosmology allows us to investigate the ability of future surveys to detect departures from GR which are compatible with currently available data; moreover, this choice allows us to easily use the Planck results as a CMB prior for the Fisher matrices we will compute.

\begin{table}[htbp]
	\centering%
	\scriptsize
	\begin{tabularx}{\linewidth}{|X|c c c c c c c|}
		\hline 
		  Parameters   & $\Omega_{{\rm m},0}$  & $\Omega_{{\rm b},0}$  & $h$  & $n_{\rm s}$  & $\sigma_8$ & $E_{11}$  & $E_{22}$ \\
		\hline 
		Fiducial   & {0.32}  & {0.05}  & {0.67}  & {0.96} & {0.822} & {0.1007}  & {0.8293}
        \\
		\hline
  \end{tabularx}
  \caption{\label{tab:cosmopars-fiducial}
  Cosmological parameters and their fiducial values in the MG parametrization used for the Fisher analysis in this work.
  }
\end{table}

In addition to the set $\theta_{\rm cosmo}$, our analysis also includes the set of free nuisance parameters $\theta_{\rm nuis}$ that enters in the theoretical expressions of the probes we consider. 
For GCsp we include the values of galaxy bias $b_{\rm g}(z)$ and the shot noise $P_{\rm shot}(z)$ in each of the $N_{\rm b}$ redshift bin we consider, while for IM we do the same with the HI bias $b_{\rm HI}(z)$, an effective bias that incorporates also the mean brightness temperature $\bar{T}_{\rm b}^2(z)$. For the angular probes we again include the galaxy bias in each bin $b_i$, together with the IA nuisance parameters for WL, i.e.\ $A_{\rm IA}$, $\beta_{\rm IA}$ and $\eta_{\rm IA}$. Therefore, the full set of free parameters $\Theta$ that enters in the Fisher matrix analysis of \autoref{eq:Fish_general} is $\Theta=\{\theta_{\rm cosmo},\theta_{\rm nuis}\}$.

\subsubsection{Non-linear settings} \label{nonlinear}
To conclude our analysis settings, we specify our choices for the small-scales limits of our theoretical predictions. For the angular probes, we choose $\ell_{\rm max} = 5000$. The choice of this scale relies on a modelling of the non-linear matter power spectrum and other associated systematics at small scales, based on the one used for instance in \cite{Euclid:2019clj}.
While N-body simulations of the non-linear evolution of perturbations are available in $\Lambda$CDM and can be used to reach such scales in the prediction of the matter power spectrum, this is not the case when dealing with the modified gravity parameterisations we use in this work. However, in order to reach $\ell_{\rm max}=5\,000$, we assume here the validity of the parameterized post-Friedmann (PPF) framework, developed in \cite{Hu:2007pj} and used extensively in \cite{Casas:2017eob}, which allows to reach the scales under examination also within our analysis.

For GCsp and IM instead, we cut the scales beyond $k_{\rm max}=0.3 \mathrm{h/Mpc}$ out of our analysis. The choice of this cut is based on the optimistic settings of \cite{Euclid:2019clj} for the spectroscopic probe and in the case of IM based on previous results by \cite{SKA_cosmo_redbook}.

\subsubsection{Radio Surveys: SKAO} \label{surveys_radio}

Our goal is to investigate the constraints that SKAO will achieve on the models of interest. 
We consider two kinds of galaxy surveys (in spectroscopy and radio-continuum), a weak lensing survey and the IM survey, all of them performed with the South-African mid-frequency array of the SKAO (see details in \cite{SKA_cosmo_redbook}).

The spectroscopic survey uses the 21-cm line from hydrogen observed in radio interferometry to detect and locate \hi-rich galaxies. The specifications we adopt for this survey are shown in \autoref{tab:desi-gcsp-specs} and we show the redshift distribution of sources and the fiducial galaxy bias for each bin in \autoref{tab:SKAOHI}. In the top panel of \autoref{fig:inversenoise-z-bins-ska} we show the inverse noise term for the this survey in red, also highlighting the five redshift bins we consider in our analysis.

The {\it continuum} survey identifies instead radio-emitting galaxies (e.g.\ star-forming or with an active radio galactic nucleus) from the reconstructed images from the interferometric data; their $z$ determination is poor, which make this survey the radio counterpart of photometric optical surveys. The reconstructed images of this survey are then used to obtain the weak lensing measurements. We report the specifications assumed for the continuum survey in \autoref{tab:ska-wl1}. 
In the lower panel of \autoref{fig:inversenoise-z-bins-ska} we plot in a solid line the expected $n(z)$ for the continuum galaxy population of SKAO \cite{SKA_cosmo_redbook} as a function of $z$, and in shaded pink rectangles we mark the boundaries of the continuum redshift bins used in our 3$\times$2pt analysis.

While SKAO will perform the previous surveys exploiting radio interferomentry, IM is the only survey that runs in single-dish mode, i.e.\ considering each dish as an independent telescope and co-adding the maps. In \autoref{tab:IMspecsNoise} we report the assumed specifications to describe the IM signal as expected from SKAO. In the top panel of \autoref{fig:inversenoise-z-bins-ska} we show the inverse noise term for the this survey and the division into eleven redshift bins in yellow.

\subsubsection{Optical Surveys: DESI and VRO}

While the focus of this paper is primarily on SKAO, we also compare and combine this with upcoming optical surveys, for which we take as an example the Dark Energy Spectroscopic Survey and the Vera C.\ Rubin Legacy Survey of Space and Time. 

\begin{figure}
	\centering
	\includegraphics[width=0.48\textwidth]{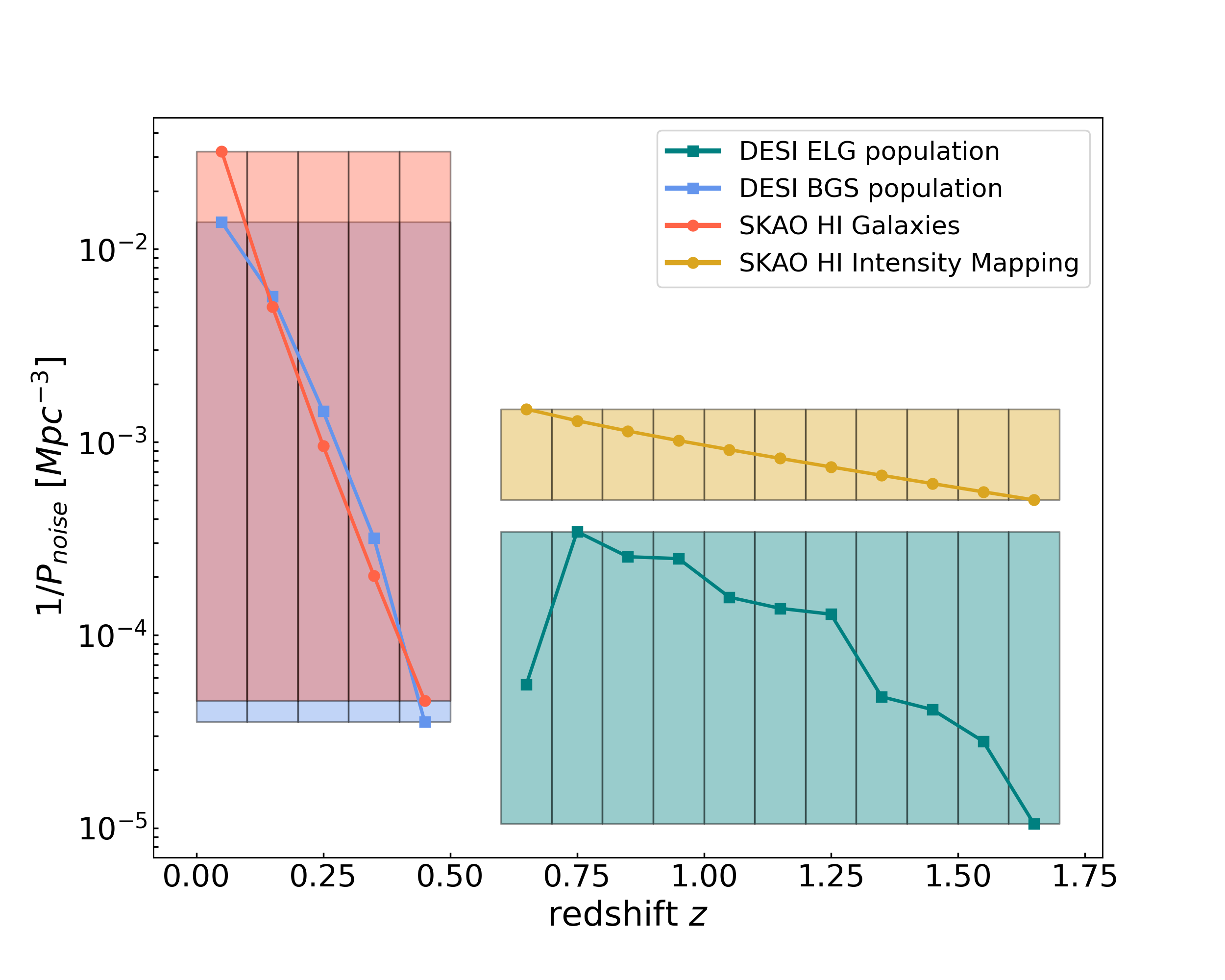}\\
	\includegraphics[width=0.48\textwidth]{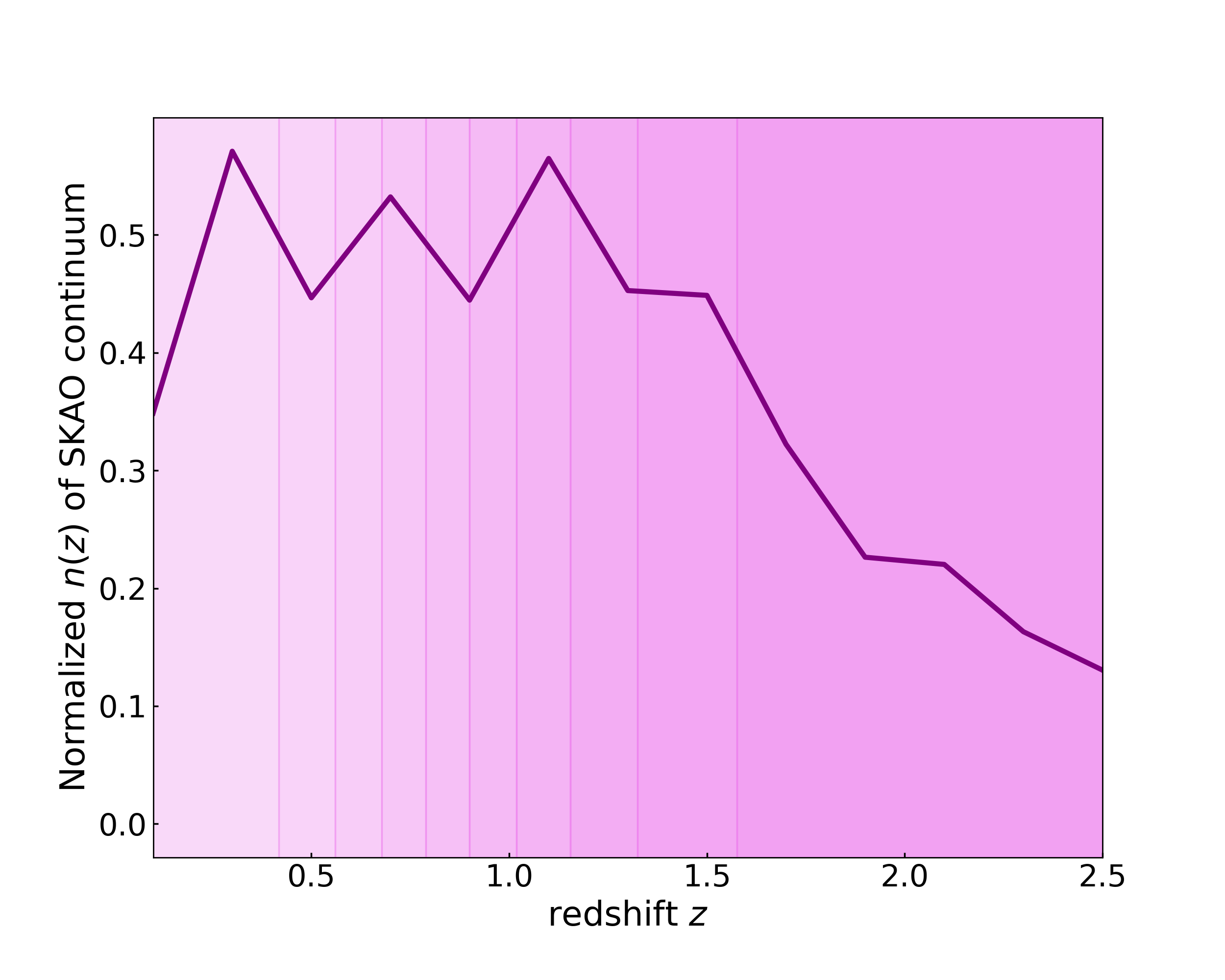}
	\caption{\textbf{Top}: Inverse noise terms ($1/P_{\rm noise}$ in units of  $[{\rm Mpc}^3]$) for SKAO GC with spectroscopic \hi\ galaxies (red, 5 bins), IM with \hi\ temperature (yellow, 11 bins), DESI H$\alpha$ BGS galaxies (light blue, 5 bins) and DESI H$\alpha$ ELG galaxies (green, 11 bins).
	\textbf{Bottom}: Normalized galaxy number density for continuum galaxies in SKAO as a function of redshift. The shaded regions correspond to the 10 photometric redshift bins. Taken from \cite{harrison_ska_2016}. 
	}
	\label{fig:inversenoise-z-bins-ska}
\end{figure}

DESI (Dark Energy Spectroscopic Survey) \cite{Vargas-Magana:2018rbb, DESI:2016fyo} is a stage-IV spectroscopic galaxy redshift survey conducted with a ground-based telescope installed in Arizona. 
While DESI will study four populations of tracers, in this work we will concentrate only on Emission Line Galaxies (ELG) between redshifts 0.7 and 1.7 and the Bright Galaxy Survey (BGS) which will cover galaxies in the range $0. < z < 0.5$. In \autoref{tab:desiELG} and \autoref{tab:desiBGS} we detail the specifications used for our forecasts, including the $N(z)$ in units of inverse volume, ${\rm Mpc}^{-3}$, and the expected galaxy bias at each redshift bin.
In the upper panel of \autoref{fig:inversenoise-z-bins-ska} we show these redshift bins and the $\bar N_i$, which corresponds to the inverse of the shot noise, for the DESI probes considered here. BGS contains 5 redshift bins, while for ELG we consider 11 redshift bins. 
For more information about the galaxy populations and the survey strategy see \cite{DESI:2016fyo} and for an overview of forecasts and the science possible with DESI, see \cite{Vargas-Magana:2018rbb, Alam:2020jdv, Font-Ribera:2013rwa}.

For our photometric analysis we consider the Vera C.\ Rubin Legacy Survey of Space and Time \cite{LSST:2008ijt, LSSTDarkEnergyScience:2012kar, LSSTDarkEnergyScience:2018jkl} (herafter VRO), which is a Stage IV galaxy survey using a ground-based telescope installed in Cerro Pachón in northern Chile.
In this work we will consider VRO for the photometric weak lensing (WLph) and clustering probes, together with their cross-correlation, i.e. the 3$\times$2pt combination (see \cite{LSSTDarkEnergyScience:2018yem}). For this survey, we model the expected galaxy number density as \cite{LSST:2008ijt, LSSTDarkEnergyScience:2012kar}
\begin{equation}\label{eq:wl-nofz}
n(z) \propto z^\beta\,\exp\left[-\left(\frac{z}{z_0}\right)^\gamma\right] \;,
\end{equation}
with $\beta=2.00$, $\gamma=1.25$ and $z_0 = 0.156/\sqrt{2}$.

In \autoref{tab:ska-wl1} we list the rest of the specifications of the VRO photometric survey used for the forecasts in this work. For more details on the survey strategy, specifications and the anticipated data products see \cite{LSST:2008ijt}.

\section{Results}\label{sec:results}

\subsection{SKAO forecasts}
\begin{figure*}
	\centering
	\includegraphics[width=0.9\linewidth]{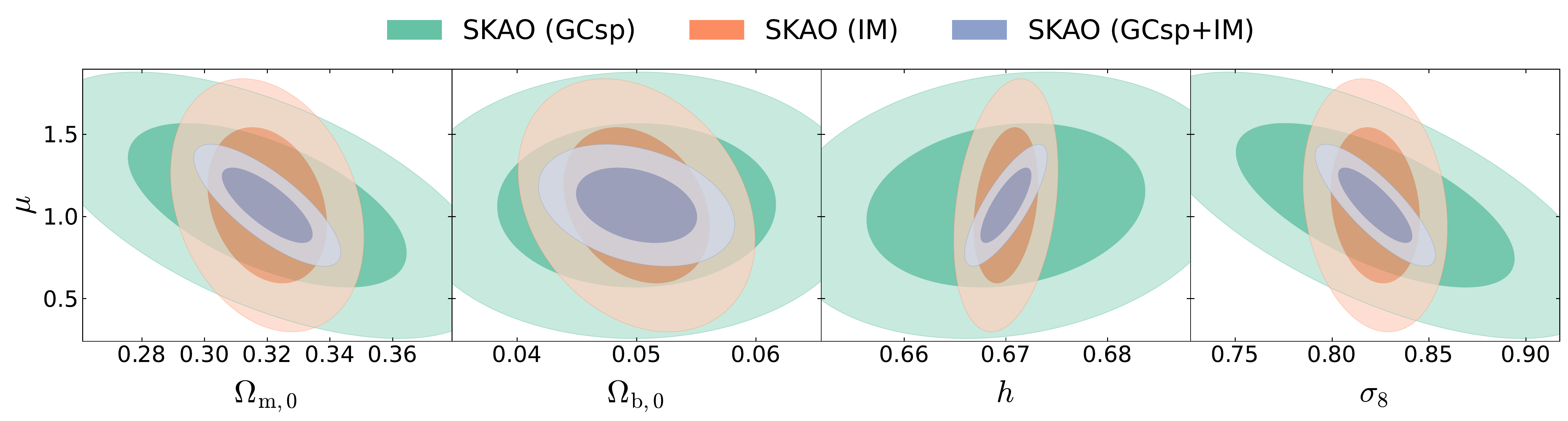}
	\caption{Fisher-matrix-marginalised forecasts on the late-time parameterisation model for SKAO. In green the GCsp probe from HI galaxies, in orange the IM probe from 21cm Intensity mapping and in violet the combination of GCsp and IM.}
	\label{fig:fisher-SKA1-Pkprobes}
\end{figure*}

For SKAO, we start from considering the two probes that 
are linked to the 3-dimensional matter power spectrum, namely spectroscopic Galaxy Clustering (GCsp) and 21-cm Intensity Mapping (IM). In \autoref{fig:fisher-SKA1-Pkprobes} we show the forecasted  1-$\sigma$ and 2-$\sigma$ confidence level contours for GCsp in green and for IM in orange, together with the combination of both in blue. Both these probes are tracing the underlying clustering of structures; therefore, we expect them to be sensitive to the MG parameter $\mu$ that affects the trajectories of massive particles. In addition, they are also sensitive to redshift space distortions and the Alcock-Paczynski effect, and therefore to parameters like $h$ and $\Omega_{{\rm b},0}$.
These two separate SKAO probes can be combined since, as shown in the top panel of \autoref{fig:inversenoise-z-bins-ska}, GCsp probes low redshifts from $ 0 < z < 0.4$ and IM probes higher redshifts $ 0.6 < z < 1.7$, such that a simple addition of their Fisher matrices is enough and we do not need to calculate their cross-correlation.
We can see in \autoref{fig:fisher-SKA1-Pkprobes} that IM dominates the constraining power for most parameters: this is due to the fact that IM probes a much larger area and a higher number of redshift bins, which allows to capture time and scale-dependent variations of the power spectrum.
In \autoref{tab:errors-late_time} we show fully marginalised constraints on different parameters, and for different probe combinations. As said, probes based on the matter power spectrum are particularly suited to constrain $\mu(z)$ and its present amplitude $\mu\equiv\mu(z=0)$, as this parameter is the one affecting the growth rate of matter perturbations (see \autoref{eq: mu_def}); we indeed find that GCsp alone can constrain its value with approximately a 31\% accuracy, which is similar to the  constraining power that IM alone is able to provide (29\%). Moreover, their combination can constrain $\mu$ at the 14\% level, a factor 2 improvement. The above holds for SKAO alone, without any priors from other experiments.
As expected, the parameter $\Sigma\equiv\Sigma(z=0)$, which mainly affects lensing, is instead not well constrained by these probes and the relative error bars amount to more than 100\% in all these cases.

\begin{figure*}
	\centering
	\includegraphics[width=0.9\linewidth]{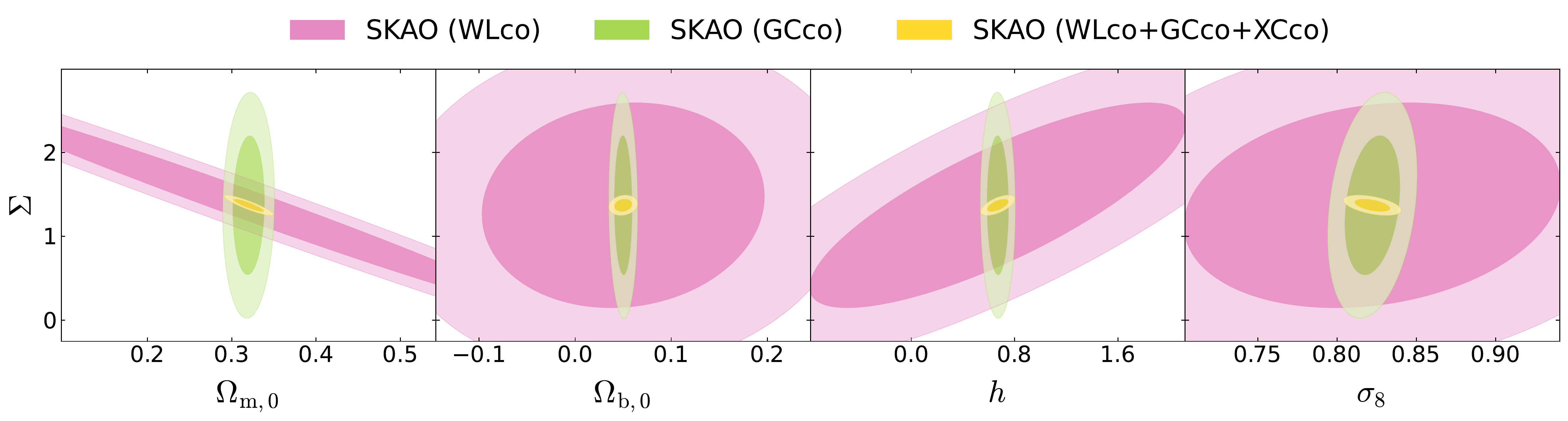}
	\caption{Fisher-matrix-marginalised forecasts on the late-time parameterisation model for SKAO. In pink the WLco probe from the continuum survey, in green the GCco probe and in yellow the combination of GCco, WLco and their cross-correlation XCco.}
	\label{fig:fisher-SKA1-Angprobes}
\end{figure*}

\begin{figure*}
	\centering
	\includegraphics[width=0.8\linewidth]{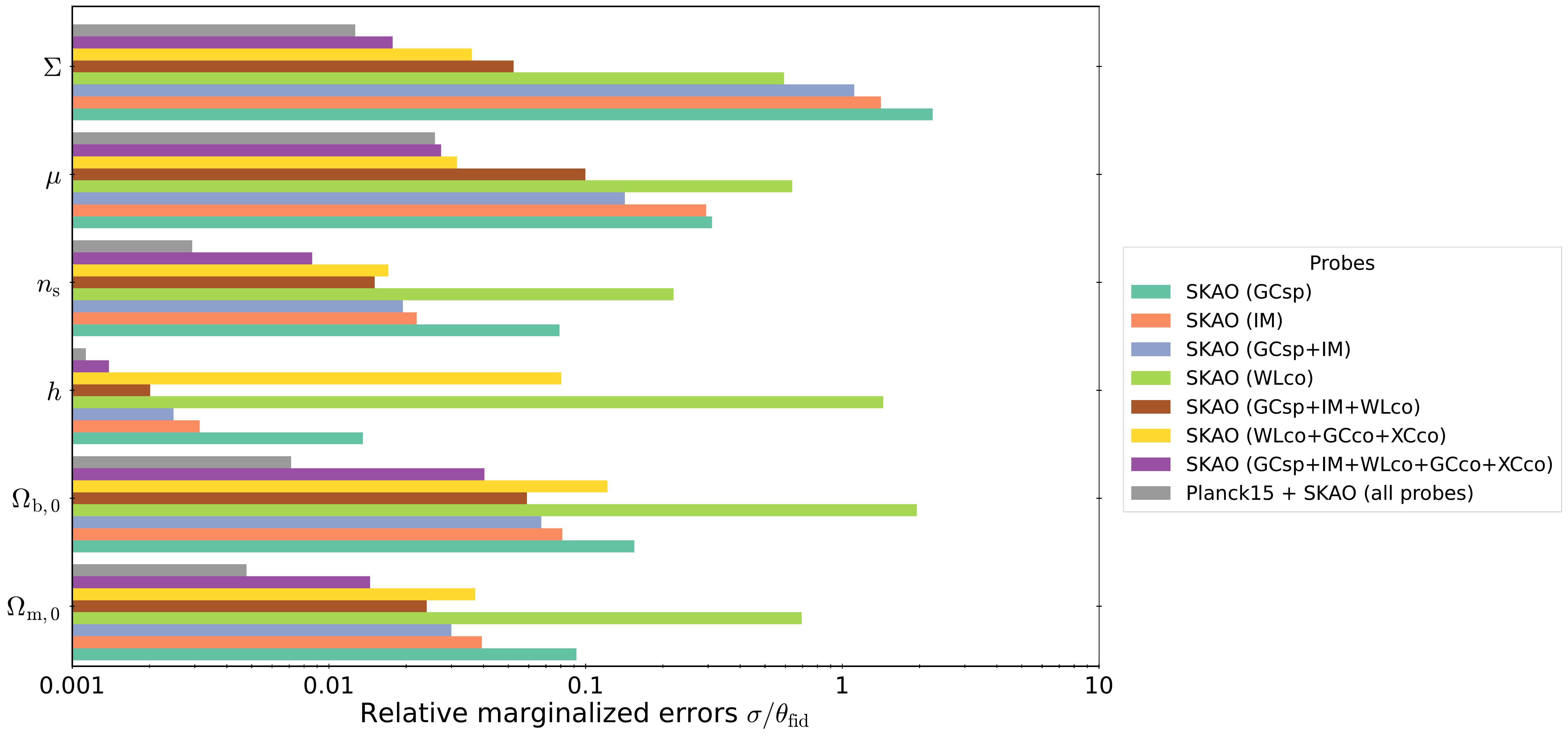}
	\caption{
	1$\sigma$ fully-marginalized constraints on the late-time parametrization model, for all SKAO probes separately and their combinations. In bluegreen GCsp, in orange IM, in blue GCsp+IM, in limegreen the WLco probe, in brown the combination GCsp+IM+WL, in yellow the 3$\times$2pt combination of continuum galaxies (GCco+WLco+XCco), in purple all SKAO probes combined and in silver the further addition of Planck15 data.}
	\label{fig:barplot-SKA1-allprobes-separate}
\end{figure*}

On the other hand, for the angular probes of SKAO, which have a low redshift resolution but a good angular resolution, we observe that the parameter $\Sigma$ is much better estimated, since it affects WLco and also the galaxy-galaxy-lensing cross-correlation (XCco).
While WLco from SKAO alone can constrain this parameter only up to about 60\% accuracy \citep[cf.][]{harrison_ska_2016,2016MNRAS.463.3686B,2017MNRAS.464.4747C}, due to its small area coverage and large shape measurement errors, the combination of WLco with continuum Galaxy Clustering (GCco) and their cross-correlation (XCco) can already constrain $\Sigma$ at the $\approx 3.6\%$ level. 
As can be seen in \autoref{fig:fisher-SKA1-Angprobes}, the large degeneracy between $\Sigma$ and $\Omega_{{\rm m},0}$ coming from WLco alone (in pink contours) is broken by the robust determination of $\Omega_{{\rm m},0}$ by GCco (in green); therefore, the full combination GCco+WLco+XCco (yellow contours) is powerful in constraining $\Sigma$.
Also, as we see in \autoref{tab:errors-late_time}, WLco+GCco+XCco is much better at constraining both $\mu$ and $\Sigma$ than GCsp+IM alone can do. When combining the GCsp and IM probes with WLco only, we obtain already a determination of $\approx 10\%$ on $\mu$ and  $\approx 5.2\%$ on $\Sigma$. An overview bar plot of how different probe combinations perform on different parameters is also shown in \autoref{fig:barplot-SKA1-allprobes-separate}. 

In \autoref{fig:fisher-SKA1-Pk-vs-Angular-separate} we compare SKAO GCsp+IM (blue), SKAO angular probes (WLco+GCco+XCco, in yellow) and their combination (purple).
We observe an excellent complementarity between probes in the planes of $\mu - \Omega_{{\rm m},0}$ and $\mu - h$; as discussed above, the combination of all angular probes constrains very well the $\mu$ parameter, but GCsp+IM are better at constraining $\Omega_{{\rm m},0}$ and $h$ on their own. The same complementarity appears with $\Sigma - \Omega_{{\rm m},0}$ for which there are no constraints on the first parameter coming from GCsp+IM; the angular probes are nearly perpendicular to GCsp+IM contours. 
Overall, the combination of angular probes and GCsp+IM indicated as SKAO$_{\rm all}$ in \autoref{tab:errors-late_time} further improves  the relative error on the $\Omega_{{\rm m},0}$ parameter by a factor 2 and the one on  $\Omega_{{\rm b},0}$ by a factor 3, with respect to angular probes alone. A small gain for the MG parameters is also present: for $\Sigma$ the error is reduced from $\sim$3.6\% to $\sim$1.8\%;  
however, for $\mu$ the effect is relatively negligible, bringing the error from 3.2\% to 2.7\%, a small difference compared to the possible systematic errors in our Fisher analysis.

So far, we have considered results relying on the information from SKAO alone. When adding Planck15 \cite{planck_collaboration_planck_2016} priors --coming from temperature, polarization and lensing reconstruction information-- to the analysis, we gain on all standard cosmological parameters, but most drastically on $\Omega_{{\rm b},0}$ and $n_{\rm s}$---by a factor four and three, respectively---which is expected due to the high sensitivity of the CMB spectra to the baryonic density and the scalar tilt. We computed these priors from the covariance matrix of the parameter posterior, inverting it to obtain a Fisher-matrix-like constrain.
Regarding MG parameters, as expected we find no gain on $\mu$, since the CMB does not probe late-time clustering (except partially from CMB-lensing). However, for $\Sigma$, there is an improvement in constraining power since the CMB can also probe the sum of the two gravitational potentials due to the ISW effect and the lensing of the CMB photons.
Indeed, when combining all probes of SKAO plus Planck15, we obtain a relative error on $\Sigma$ of 1.3\%, which would already be enough to test, and potentially rule out, most of the compelling models of modified gravity.

Another notable result from these forecasts is the constraining power that SKAO will have on the Hubble parameter $h$ just from GCsp and IM alone.
When considering just the GCsp probe, SKAO will determine $h$ with a 1.4\% accuracy, which then gets improved by roughly a factor 7 to 0.2\% when adding the information coming from IM.
While the angular probes on their own will not be able to determine the Hubble parameter with great accuracy, they do contain some information due to the sensitivity of the GCco probe, that helps to break some degeneracies; angular probes and GCsp and IM improves by almost a factor 2 the constrain on $h$ to a relative $1\sigma$ statistical error of 0.1\%. The combination with Planck at this stage does not add any further significant constraint on the Hubble parameter.
These results are all listed in \autoref{tab:errors-late_time} and displayed as a bar plot in \autoref{fig:barplot-SKA1-allprobes-separate}.
 
\begin{figure*}
	\centering
	\includegraphics[width=0.6\linewidth]{new-plots-final/SKAO_angprobes+pkprobes_MuSigma.pdf}
	\caption{Fisher-matrix-marginalised forecasts on the late-time parametrization model, for SKAO. We show separately the combination of GCsp+IM (in blue), the combination GCco+WLco+XCco (in yellow), and the full combination of these in purple. The parameters shown here are $\Omega_{{\rm m},0}$, $h$, $\sigma_8$ and the MG parameters $\mu$ and $\Sigma$.}
	\label{fig:fisher-SKA1-Pk-vs-Angular-separate}
\end{figure*}

\begin{table}[htbp]
	\centering%
	\scriptsize
	\begin{tabularx}\linewidth{|l|cccccc|}
		\hline 
		SKAO & $\Omega_{{\rm m},0}$  & $\Omega_{{\rm b},0}$  & $h$  & $n_{\rm s}$  & $\mu$    & $\Sigma$ 
		\tabularnewline
		\hline 
		\multicolumn{1}{|l|} {\it Fiducial}   & {0.32}  & {0.05}  & {0.67}  & {0.96} & {1.07}  & {1.37}  \tabularnewline
		\hline
		\hline
		 GCsp
	    &9.2\%   &15.5\%    &1.4\%    &7.9\%    &31.0\%    &224\%  \tabularnewline
		 IM  
		&3.9\%	&8.1\%	&0.3\%	&2.2\%	&29\%	 &141\% 
		\tabularnewline
		GCsp+IM
		&3.0\%	&6.7\%	&0.2\%	&1.9\%	&14\%	&111\%  
		\tabularnewline
		\hline
		 WLco 
		&69.4\%	&194\%	&144\%	&22\%	&63\%	&59\%  
		\tabularnewline
		 GCsp+IM+WLco 
		&2.4\%	&5.9\%	&0.2\%	&1.5\%	&10\%	&5.2\% 
		\tabularnewline
		 WLco+GCco+XCco    
		&3.7\%	&12.2\%	&8.0\%	&1.7\%	&3.2\%	&3.6\%  
		\tabularnewline
		\Tstrut SKAO$_{\rm all}$ 
	    &1.4\%	&4.0\%	&0.1\%	&0.9\%	&2.7\%	&1.8\%  
	    \tabularnewline
		\hline
		\hline
		 SKAO$_{\rm all}$+Planck15
		 &0.5\%	&0.7\%	&0.1\%	&0.3\%	&2.6\%	&1.3\%  \tabularnewline
		 \hline
		  
	\end{tabularx}
	\caption{\label{tab:errors-late_time}
		1$\sigma$
		fully marginalized percentage errors on the cosmological parameters in the late-time parameterisation of Modified Gravity for the different SKAO probes.  With the label SKAO$_{\rm all}$  we refer to the combination of GCsp+IM+GCco+WLco+XCco. In the last line we show the improvement in constraints when including a Planck15 prior. Here the parameters $\mu$ and $\Sigma$ are the present day values of the MG functions $\mu(z)$ and $\Sigma(z)$.}
\end{table}

\subsection{Combination of SKAO and next-generation optical probes}\label{sec:optcombradio}
In this section we look at the complementarity between SKAO and optical galaxy survey experiments such as the Vera C.\ Rubin Observatory (VRO) and the Dark Energy Spectroscopic Instrument (DESI), which will perform wide-field galaxy surveys, using photometric and spectroscopic probes, respectively. 
In the case in which the probes can be considered independent as a first approximation, we perform this combination by simply adding their respective Fisher matrices together, as it is the case when combining photometric probes and spectroscopic probes together, or when combining spectroscopic probes in different redshift ranges. This is what we do for the combination of the angular continuum probes of SKAO plus VRO or DESI.

\begin{figure*}
	\centering
	\includegraphics[width=0.8\linewidth]{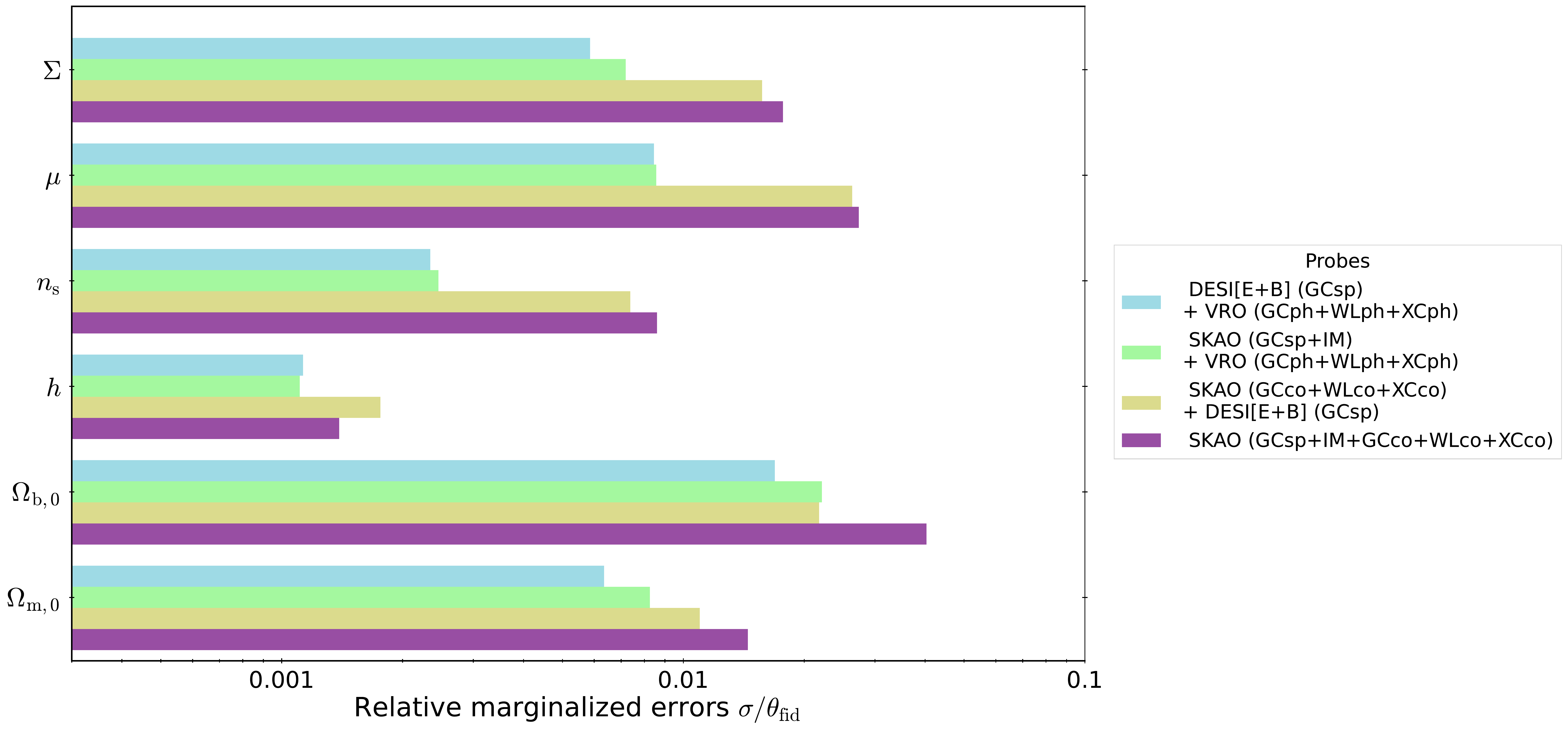}
	\caption{
	1$\sigma$ fully-marginalized constraints on the late-time parametrization model, for different surveys and combinations. In purple all the combined probes of SKAO (GCsp+IM+GCco+WLco+XCco), in cyan the full combination of DESI and VRO (spectroscopic and photometric), in bright green the combination of GCsp+IM of SKAO combined with the angular probes of VRO (GCph+WLph+XCph), in teal green the continuum observables of SKAO (GCco+WLco+XCco) combined with the spectroscopic probes of DESI ELG and BGS (GCsp).
	When combining photometric probes from a stage-IV survey like VRO with the redshift-accurate probes of SKAO, the constraints are very powerful and very similar to VRO+DESI only, while offering a different degeneracy breaking of systematics. }
	\label{fig:barplot-SKA1-DESI-Rubin-separate}
\end{figure*}

\begin{figure*}
	\centering
	\includegraphics[width=0.7\linewidth]{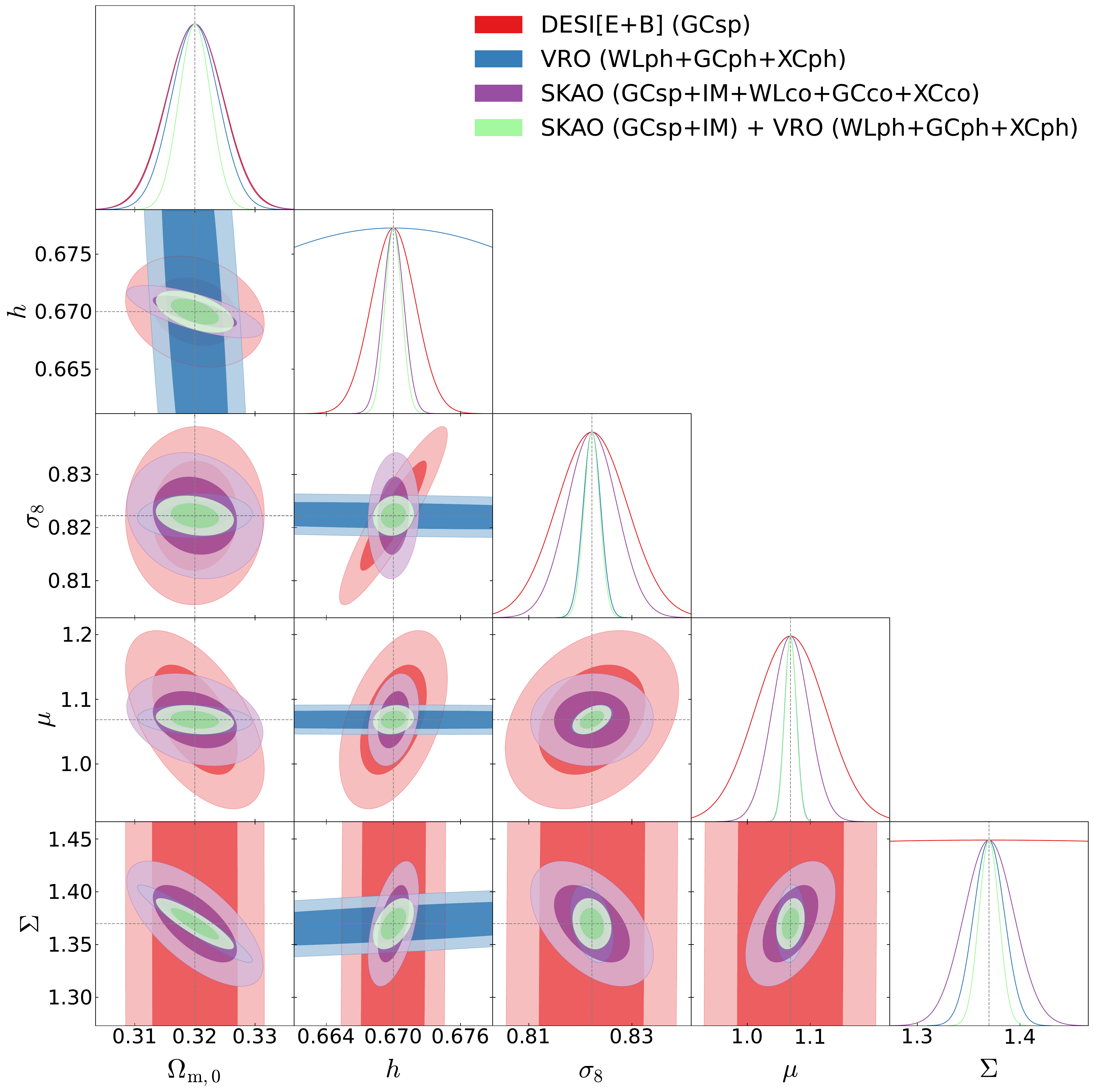}
	\caption{
	Fisher-matrix-marginalised forecasts on the late-time parameterisation model. We show ${\rm DESI}$ (the GCsp probe of the DESI ELG +  BGS samples) in red, VRO (GCph+WLph+XCph) in blue, in purple all SKAO probes combined together and in green the combination of the photometric VRO + the IM and GCsp surveys of SKAO.}
	\label{fig:fisher-DESI-Rubin}
\end{figure*}

In \autoref{fig:barplot-SKA1-DESI-Rubin-separate} and in \autoref{tab:errors-late_time-combined} we show the 1$\sigma$ fully-marginalized constraints on the late-time parameterization model, for different surveys and combinations of probes. In purple all the combined probes of SKAO (GCsp+IM+GCco+WLco+XCco) discussed so far, in cyan the full combination of DESI and VRO (spectroscopic plus photometric), in bright green the combination of SKAO GCsp+IM with photometric probes of VRO (GCph+WLph+XCph), in teal green the continuum observables of SKAO (GCco+WLco+XCco) combined with the spectroscopic probes of DESI ELG and BGS (GCsp).
The full constraining power of the combination of SKAO probes is competitive in some of the cosmological parameters with a full combination of Stage-IV galaxy surveys, such as DESI+VRO (cyan), as it is the case of the Hubble parameter $h$: from \autoref{tab:errors-late_time-combined} this is already very well constrained by the GCsp+IM of SKAO alone at the 0.2\% level and, when considering all SKAO probes together, the constraint decreases to 0.1\%.
For the parameters of interest in modified gravity, namely $\mu$ and $\Sigma$, 
the best constraints are obtained when using the combination of all angular probes (GCph+WLph+XCph) information from VRO, either combined with GCsp and IM from SKAO (bright green bar in \autoref{fig:barplot-SKA1-DESI-Rubin-separate}) or combined with
DESI (cyan bar in \autoref{fig:barplot-SKA1-DESI-Rubin-separate}). 
In the former case, the $1\sigma$ constraints on $\mu$ and $\Sigma$ are 0.9\% and 0.7\%, respectively, while in the latter case they are slightly improved at 0.8\% and 0.6\%, respectively. These results are also shown in \autoref{fig:fisher-DESI-Rubin}, where we plot the forecasted $1$- and $2$-$\sigma$ fully-marginalized contours on the late-time parametrization, from the spectroscopic DESI probe (red), the photometric probe of VRO (blue), the full combination of SKAO probes (purple) and the combination of VRO photometric with GCsp+IM from SKAO (bright green).

In summary, this means that when combining photometric probes from a stage-IV survey like VRO with the GCsp+IM of SKAO, the constraints on modified gravity are very similar to VRO+DESI only, with the advantage that they could offer a different breaking of systematic effects.

\begin{table}[htbp]
	\centering%
	\scriptsize
	\begin{tabular}{|l|cccccc|}
		\hline 
		SKAO & $\Omega_{{\rm m},0}$  & $\Omega_{{\rm b},0}$  & $h$  & $n_{\rm s}$  & $\mu$    & $\Sigma$ 
		\\
		\hline 
		\multicolumn{1}{|l|} {\it Fiducial}   & {0.32}  & {0.05}  & {0.67}  & {0.96} & {1.07}  & {1.37}  \\
		\hline
		\hline
		SKAO$_{\rm all}$  & \multirow{1}{*}{1.4\%}	& \multirow{1}{*}{4.0\%}	& \multirow{1}{*}{0.1\%}	& \multirow{1}{*}{0.9\%}	& \multirow{1}{*}{2.7\%}	& \multirow{1}{*}{1.8\%} \\
		\hline
		DESI$_{\rm E+B}$ (GCsp) & \multirow{2}{*}{1.1\%}	& \multirow{ 2}{*}{2.2\%}	& \multirow{ 2}{*}{0.2\%}	& \multirow{2}{*}{0.7\%}	& \multirow{ 2}{*}{2.6\%}	& \multirow{ 2}{*}{1.6\%} \\
		+ SKAO (angular) & & & & & &  \\
		\hline
        SKAO (GCsp+IM) & \multirow{ 2}{*}{0.8\%}	& \multirow{ 2}{*}{2.2\%}	& \multirow{ 2}{*}{0.1\%}	& \multirow{ 2}{*}{0.2\%}	& \multirow{ 2}{*}{0.9\%}	& \multirow{ 2}{*}{0.7\%} \\
		+ VRO (angular) & & & & & &  \\		
		\hline
	   DESI$_{\rm E+B}$ (GCsp) & \multirow{ 2}{*}{0.6\%}	& \multirow{ 2}{*}{1.7\%}	& \multirow{ 2}{*}{0.1\%}	& \multirow{ 2}{*}{0.2\%}	& \multirow{ 2}{*}{0.8\%}	& \multirow{ 2}{*}{0.6\%} \\
		+ VRO (angular) & & & & & &  \\
		\hline
		  
	\end{tabular}
	\caption{\label{tab:errors-late_time-combined}
		1$\sigma$
		fully marginalized errors on the cosmological parameters in the late-time parameterisation of Modified Gravity for a different combination of probes. We refer to the combination GCsp+IM+GCco+WLco+XCco with SKAO$_{\rm all}$, to GCco+WLco+XCco with SKAO (angular) and to GCph+WLph+XCph with VRO (angular).}
\end{table}

\subsection{IM and GCsp cross-correlation}\label{sec:DESIxSKAO}

In the previous section we have investigated the information contained in the combination of radio and optical surveys. In order to perform this combination we had to avoid combining probes overlapping in redshift, as this would lead to double counting information if cross-correlations are neglected. For such a reason, in \autoref{sec:optcombradio} we have combined the GCsp from DESI with the angular probes of SKAO, but avoided the combination with GCsp and IM from SKAO as the latter overlaps in redshift with DESI ELG galaxies (see \autoref{fig:inversenoise-z-bins-ska}).

Nevertheless, avoiding such a combination is potentially depriving us of additional constraining power. Therefore, we explore here the impact of the full combination SKAO (GCsp, IM and angular probes) with DESI (GCsp from ELG galaxies), including the corss-correlation between IM and DESI galaxies using \autoref{eq:GC_Cov_IMg} with the power spectrum of \autoref{eq:PIMg-cross}.

\begin{figure*}
	\centering
	\includegraphics[width=0.9\linewidth]{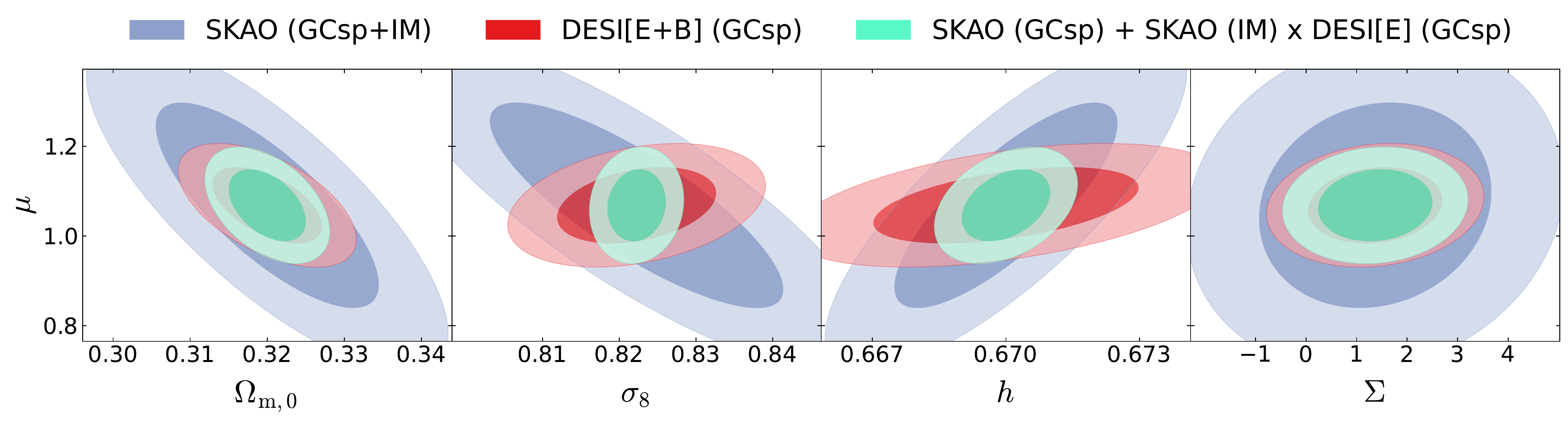}
	\caption{Fisher-matrix-marginalised forecasts on the late-time parameterisation model for the combination of different observables with SKAO and DESI. In blue the 21cm IM probe from SKAO combined with the HI GCsp probe of the same experiment, in red GCsp from the combination of two DESI surveys, ELG and BGS, labeled as DESI[E+B] here. Finally, in bright green, the combination of DESI ELG spectroscopic galaxies with SKAO 21cm IM, including its cross-correlation in the redshift range $0.65<z<1.75$, as detailed in \autoref{eq:GC_Cov_IMg} plus the SKAO HI GCsp, which probes low redshifts $z<0.5$.
    The use of cross-correlation between IM and GCsp improves considerably the constraints as opposed to a GCsp survey alone, especially for the parameters $h$ and $\sigma_8$.}\label{fig:IMxGCsp}
\end{figure*}

In \autoref{fig:IMxGCsp} we show the impact of adding this cross correlation. We show in blue the constraints obtained from the GCsp+IM of SKAO, and in red the constraints obtained from DESI ELG + BGS galaxies, highlighting the complementarity of the information coming from the two surveys. In bright green we show instead the combination of SKAO GCsp+IM with DESI ELG, where the cross-correlation between the latter two has been included and DESI BGS galaxies have been removed (as they completely overlap in redshift with SKAO GCsp). The results show how including this extra information significantly improves the constraints: while the bounds on $\mu$ are very similar to those achievable with DESI ELG + BGS, in $\Sigma$ we can see an improvement of about 15\%. On the other hand the biggest gain is observed on standard parameters, where the constraints on $\sigma_8$ and $h$ improve by a factor $2.7$ and $3.0$ respectively.

\begin{figure*}
	\centering
	\includegraphics[width=0.8\linewidth]{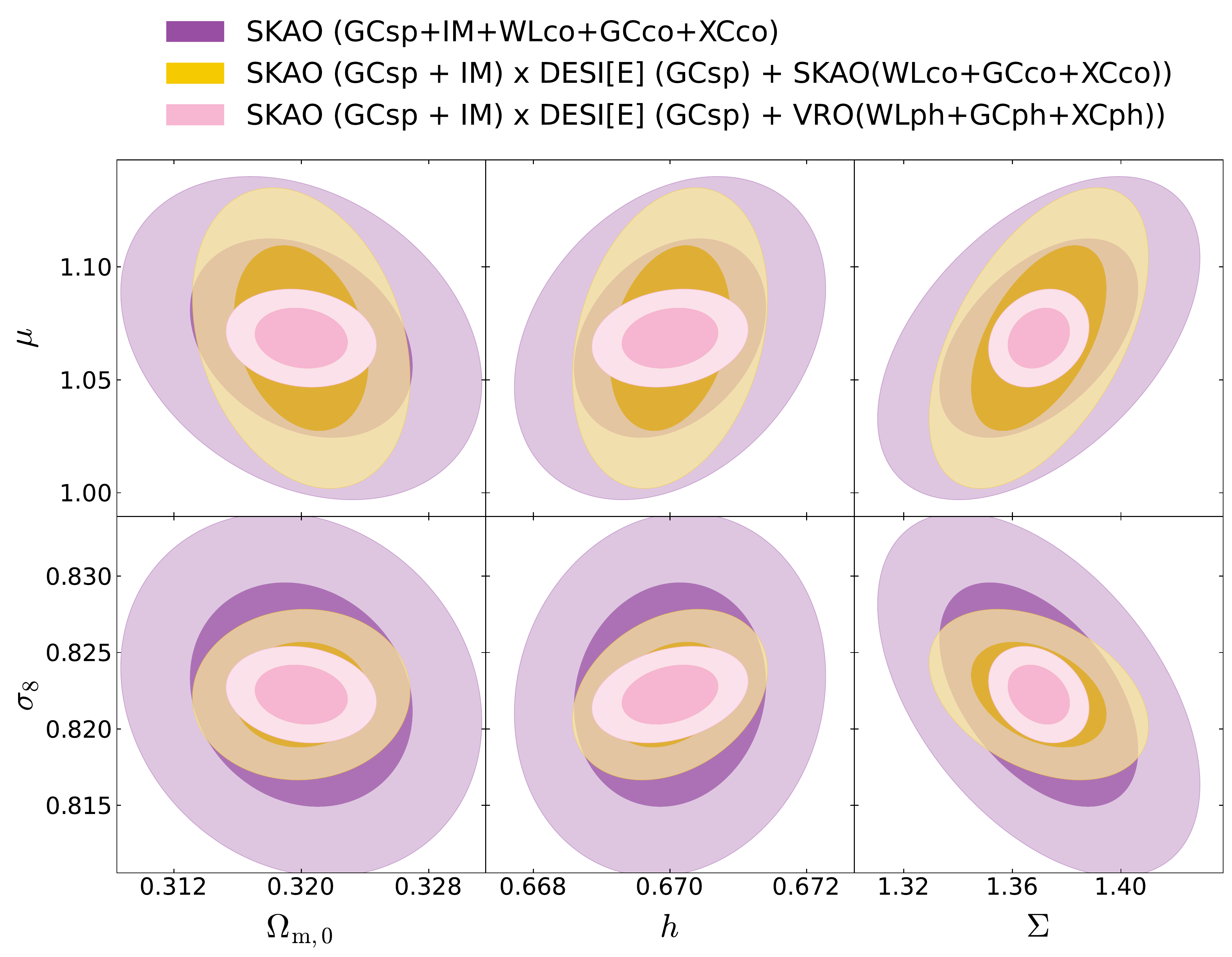}
	\caption{
	Fisher-matrix-marginalised forecasts on the late-time parameterisation model for different combinations of probes. We show in purple all SKAO probes combined together on their own, in yellow we show the marginalised contours when also the SKAO IM x DESI ELG cross-correlation is added to the combination. Finally, in pink, we show our most-constraining combination, consisting of SKAO GCsp, SKAO IM, DESI ELG and their cross-correlation, combined with all the photometric probes of VRO (which also include their own cross-correlation). This last combination of SKAO, DESI ELG and VRO in pink yields a factor 3 and 4 improvement in the determination of $\mu$ and $\Sigma$, respectively, over the previous combination, in yellow, of just SKAO and DESI ELG.}
	\label{fig:fisher-SKA_all-x-DESI}
\end{figure*}

In \autoref{fig:fisher-SKA_all-x-DESI} we show the comparison between the constraints achievable combining all SKAO probes (GCsp+IM+GCco+WLco+XCco) with the case in which the cross-correlation between IM and DESI ELG is added to the combination. 
While the improvement is not striking on $\mu$, we can see how the constraints on $\Sigma$, $\Omega_{\rm m,0}$, $h$ and $\sigma_8$ are all significantly improved, respectively by a factor $1.5$, $1.6$, $1.6$, $2.1$.

In addition to these combinations, we show with the pink contours of \autoref{fig:fisher-SKA_all-x-DESI} our best constraints for all cosmological parameters, obtained when combining photometric probes from VRO, IM and GCsp from SKAO, and including the IMxGCsp cross-correlation between DESI ELG and SKAO. This is one of our major results: cross-correlation between optical and radio surveys significantly improves constraints on both standard and MG cosmological parameters, with respect to all SKAO probes.

In \autoref{tab:errors-best}, we show that comparing with the full combination of optical probes, the cross-correlation improves slightly the constraints on the modified gravity parameters, and further offers a much tighter determination of the Hubble parameter $h$. Moreover, its main advantage would be to break several degeneracies both of systematics and nuisance parameters, which would provide a more robust cosmological parameter estimation.

\begin{table}[htbp]
	\centering%
	\scriptsize
	\begin{tabularx}\linewidth{lcccccc}
		\hline 
		& $\Omega_{{\rm m},0}$  & $\Omega_{{\rm b},0}$  & $h$  & $n_{\rm s}$  & $\mu$    & $\Sigma$ 
		\tabularnewline
		\hline 
		 & $0.32$  & $0.05$ & $0.67$ & $0.96$ & $1.07$ & $1.37$ \\
	    \hline
       \hline
		DESI$_{\rm E+B}$ (GCsp) & \multirow{ 2}{*}{0.64\%}	& \multirow{ 2}{*}{1.69\%}	& \multirow{ 2}{*}{0.11\%}	& \multirow{ 2}{*}{0.23\%}	& \multirow{ 2}{*}{0.84\%}	& \multirow{ 2}{*}{0.59\%} \tabularnewline 
		+ VRO (angular) & & & & & &  \tabularnewline
           \hline
            SKAO(angular)  & \multirow{ 3}{*}{0.87\%}	& \multirow{ 3}{*}{1.91\%}	& \multirow{ 3}{*}{0.09\%}	& \multirow{ 3}{*}{0.57\%}	& \multirow{ 3}{*}{2.55\%}	& \multirow{ 3}{*}{1.2\%} \tabularnewline 
		 + SKAO (GCsp)   & & & & & &  \tabularnewline
		+ SKAO x DESI$_{\rm E}$  & & & & & &  \tabularnewline
		\hline
		 SKAO (GCsp+IM) & \multirow{ 3}{*}{0.6\%}	& \multirow{ 3}{*}{1.51\%}	& \multirow{ 3}{*}{0.07\%}	& \multirow{ 3}{*}{0.23\%}	& \multirow{ 3}{*}{0.83\%}	& \multirow{ 3}{*}{0.55\%} \tabularnewline 
		+ VRO (angular) & & & & & &  \tabularnewline
		+ SKAO x DESI$_{\rm E}$  & & & & & &  \tabularnewline
		\hline
		  
	\end{tabularx}
	\caption{\label{tab:errors-best}
		Same as in \autoref{tab:errors-late_time-combined} but highlighting here the effect of the cross-correlation between SKAO and DESI ELG.
            The constraints from GCsp + angular probes for Stage-IV surveys are already very tight on their own (third row). If one replaces the angular probes of VRO by the ones of SKAO and uses instead the 21cm and spectroscopic probes of SKAO correlated with DESI ELG (fourth row), one finds only slightly degraded constraints, with the advantage of having completely different systematics. This case corresponds to the yellow contours of \autoref{fig:fisher-SKA_all-x-DESI}. If one then replaces in that combination the angular probes of SKAO with the ones by VRO (corresponding to the pink contours in \autoref{fig:fisher-SKA_all-x-DESI}), the constraints become again tighter and even marginally better than the ones in the VRO+DESI case alone. 
            }
\end{table}

\section{Conclusions}\label{sec:conclusions}

The SKA Observatory will be the biggest radio telescope ever built. When it will start observing, cosmology in the radio band will reach a new status, becoming as competitive as the more traditional observational approaches (e.g.\ optical galaxy surveys, Lyman-alpha forest transmission measurements, and the Cosmic Microwave Background).
In this work, we focused on what such novel radio surveys can tell us about modifications of gravity and on how their synergy and cross-correlation with optical surveys can improve constraints. With a Fisher formalism approach, we derived the constraints on beyond-$\Lambda$CDM gravity theories that we will achieve with the specifications of the SKAO Mid radio telescope. In particular, we considered the $\mu$ and $\Sigma$ functions, whose departure from unity are signs of deviations from Einstein's General Relativity: $\mu$ measures a modification in the growth of perturbations, while $\Sigma$ measures a modification in the lensing amplitude.
We considered four probes carried out with SKAO surveys:  1) a clustering GCco and 2) weak lensing WLco survey performed with galaxies detected in the radio-continuum ---whose angular information outperforms the radial; 3) a spectroscopic galaxy clustering survey GCsp via \hi\,  galaxies and 4) \hi\, intensity mapping, IM.
The spectroscopic \hi\,  galaxies retain optimal angular and radial information but only with a small observed area at low redshift ($z\lesssim1$). On the other hand, the \hi\,  intensity maps share the good spectral information from the \hi-emitted 21-cm line and contain the angular information to reach a larger area and depth in the radial direction. 

In order to explore the synergies of SKAO with optical surveys, we have further considered three probes: 
1) a clustering GCph and 2) weak lensing WLph survey from photometric observations of galaxies; 3) a clustering GCsp obtained through a spectroscopic galaxy survey.

We here summarize our main results. We focus on the modified gravity parameters $\mu$ and $\Sigma$ and the Hubble parameter $h$ since it could also have exciting repercussions for addressing the current tensions in cosmological data. 
Results for SKAO are summarised in \autoref{tab:errors-late_time}. For the GCsp+IM of SKAO alone, the constraining power on these three parameters amounts to 14\%, 111\% (i.e. no constraint on $\Sigma$) and 0.2\%, respectively.
For the probes coming from the continuum survey and their cross correlation (WLco+GCco+XCco) the forecast error on $\mu$ improves to 3.2\%, the one on $\Sigma$ to 3.6\%, while the one on $h$ is, as expected, less constraining at 8\%.
When combining GCco+WLco+XCco with GCsp and IM (what we refer to as SKA$_{\rm all}$), the constraints improve to 2.7\%, 1.8\% and 0.1\% for  $\mu$,  $\Sigma$ and $h$, respectively.
We found that adding Planck priors on top of SKA$_{\rm all}$ does not help to constrain any of these parameters better, however, as expected, it reduces the statistical errors on other parameters, such as $\Omega_{{\rm m},0}$ and $\Omega_{{\rm b},0}$.

Furthermore, we decided to study the combination of SKAO probes with next-generation galaxy surveys such as DESI and VRO (\autoref{fig:fisher-DESI-Rubin}) and we found that the combination of SKAO (GCsp+IM) with VRO yields the most constraining power, being almost as good as a combination of DESI+VRO (comparison in  \autoref{tab:errors-late_time-combined}). This is due to the good redshift resolution and wide range being exploited in the SKAO combination of spectroscopic galaxy clustering and intensity mapping. When combined with a photometric 3$\times$2pt probe such as VRO, it yields very tight constraints on both the $\mu$ parameter and the $\Sigma$ parameter. The combination of SKAO GCsp+IM with VRO, improves the constraints on $\mu$ by a factor 3, while the constraint on $\Sigma$ is a factor 2.6 better than using the full combination of SKAO probes on their own. On the Hubble parameter $h$, the constraint is just slightly better and stays at the 0.1\% level.

While the synergies explored above are already extremely powerful, we demonstrated in this paper that further constraining power can be obtained considering cross-correlations. In particular, IM can be exploited  alongside other surveys obtaining strong constraints on beyond \lcdm models.\footnote{While completing this paper, \cite{Scelfo:2022lsx} appeared on the ArXiv independently. While \cite{Scelfo:2022lsx} investigates synergies with gravitational waves, we rather focus on cross-correlation with optical surveys. Choices in scale cuts and non-linear assumptions are also different; we discuss our prescriptions in \autoref{nonlinear}, including the PPF formalism allowing us to reach smaller scales.}

In this work, given our interest in the synergies between radio and optical surveys, we considered the cross-correlation between the IM survey of SKAO and the ELG galaxies of DESI. We have shown in \autoref{sec:DESIxSKAO}, \autoref{fig:fisher-SKA_all-x-DESI} how this improves our constraints, and how the most constraining combination of probes we found in this work needs to consider such a correlation (see \autoref{tab:errors-best}).  While we have included the cross-correlation within GCsp and IM or within angular probes, we have neglected the cross-correlation of the angular probes with the GCsp and IM probes, which we leave for a future study.

We have assumed the deviations from GR to be time-dependent only, but in a further work a more in-depth study should be done, in which these parameters are also allowed to have a scale-dependence. However, in that case, the theoretical modelling of the non-linear power spectrum becomes more complicated and certain scale-cuts would make the determination of these parameters rather difficult.

While the combination of next-generation photometric and spectroscopic optical surveys yields better constraints than the SKAO probes only (\autoref{fig:barplot-SKA1-DESI-Rubin-separate}, cyan versus purple), the combination of SKAO with optical probes (either with DESI or VRO) is almost as competitive as optical surveys alone, with the advantage that the systematics and the noise terms for radio and optical probes are very different and an interesting breaking of degeneracies is at play.
The next decade of radio and optical surveys will be definitely revolutionary in terms of the information content that we will be able to obtain from the large scale structures of the Universe, allowing us to constrain models that have never been tested before, provided we can overcome certain still unsolved challenges in theoretical modelling, especially in the non-linear regime.

\section*{CRediT authorship contribution statement}
\textbf{Santiago Casas:} Conceptualization; Methodology; Software; Validation; Visualization; Writing - original draft; Writing - review \& editing. \textbf{Isabella P. Carucci:} Methodology; Resources; Writing - original draft; Writing - review \& editing. \textbf{Valeria Pettorino:} Conceptualization; Resources; Methodology; Writing - original draft; Writing - review \& editing; Supervision; Funding acquisition. \textbf{Stefano Camera:} Conceptualization; Resources; Methodology; Writing - original draft; Writing - review \& editing. \textbf{Matteo Martinelli:} Conceptualization; Methodology; Software; Writing - original draft; Writing - review \& editing.

\section*{Declaration of competing interest}

The authors declare that they have no known competing financial interests or personal relationships that could have appeared to influence the work reported in this paper.

\section*{Acknowledgements}
Sa.C. acknowledge support of CNRS/CNES postdoctoral fellowship grant at the start and during the preparation of this work.
St.C.\ and I.P.C.\ acknowledge support from the `Departments of Excellence 2018-2022' Grant (L.\ 232/2016) awarded by the Italian Ministry of University and Research (\textsc{miur}) and from the `Ministero degli Affari Esteri della Cooperazione Internazionale (\textsc{maeci}) -- Direzione Generale per la Promozione del Sistema Paese Progetto di Grande Rilevanza ZA18GR02. M.M.\ acknowledges funding by the Agenzia Spaziale Italiana (\textsc{asi}) under agreement no. 2018-23-HH.0 and support from INFN/Euclid Sezione di Roma.

\newpage 

\appendix

\section{Spectroscopic survey specifications}\label{sec:GCspspecs}
\begin{table}[htbp]
	\centering
	\begin{tabular}{cccccc}
		 \multicolumn{3}{c}{GCsp surveys} \\
		\hline
		\hline 
         & $\sigma_{z}^{\rm sp}$ & $N_{\rm b}^{\rm sp}$ \\
		\hline
         DESI BLG & $0.001$ & $5$ \\
         DESI ELG & $0.001$ & $10$ \\
         SKAO HI gal.\ & $0.001$ & $10$ \\
		\hline 
	\end{tabular}
	\caption{ \label{tab:desi-gcsp-specs} Specifications for the GCsp surveys considered in this work. Spectroscopic redshift error $\sigma_{z}^{\rm sp}$ and number of independent redshift bins $N_{\rm b}^{\rm sp}$.}
\end{table}

\begin{table}[htbp]
	\centering
 \begin{tabular}{cccc}
	\multicolumn{4}{c}{DESI BGS survey}\\
    \hline
    \hline 
    $z_{\rm min}$ & $z_{\rm max}$ & $\bar N_i\,[\mathrm{Mpc}^{-3}]$ & $b_{\rm g}$     \\
    \hline
    0.0                         & 0.1                     &  $1.38 \times 10^{-2}$  & 1.364 \\
    0.1                         & 0.2                     &  $5.68 \times 10^{-3}$   & 1.388 \\
    0.2                         & 0.3                     &  $1.44 \times 10^{-3}$    & 1.410 \\
    0.3                         & 0.4                     &  $3.18 \times 10^{-4}$      & 1.432  \\
    0.4                         & 0.5                     &  $3.54 \times 10^{-5}$      & 1.457 \\
    \hline 
    \end{tabular}
    \caption{Specifications for the BGS galaxy sample of DESI, showing for each redshift bin the range, the expected number density of samples and the fiducial value of the galaxy bias.
    The $\bar N_i$ are computed from tables of $n(z)$ by the DESI collaboration \cite{desi_collaboration_desi_2016}, by integrating the differential number density in the volume $\de V(z)$ of the redshift bin, evaluated at our fiducial cosmology.
    }\label{tab:desiBGS}
\end{table}

\begin{table}[htbp]
	\centering
   \begin{tabular}{cccc}
	\multicolumn{4}{c}{DESI ELG survey}\\
    \hline 
    $z_{\rm min}$ & $z_{\rm max}$ & $\bar N_i\,[\mathrm{Mpc}^{-3}]$  &$b_{\rm g}$     \\
    \hline
    \hline
    0.7 & 0.8 & $3.43 \times 10^{-4}$   &  1.048 \\
    0.8 & 0.9 & $2.54 \times 10^{-4}$   &  1.078 \\
    0.9 & 1.0 & $2.49 \times 10^{-4}$  & 1.110 \\
    1.0 & 1.1 & $1.57 \times 10^{-4}$ & 1.142 \\
    1.1 & 1.2 & $1.37 \times 10^{-4}$  &  1.176 \\
    1.2 & 1.3 & $1.28 \times 10^{-4}$  &  1.211 \\
    1.3 & 1.4 & $4.78 \times 10^{-5}$   &   1.247 \\
    1.4 & 1.5 & $4.11 \times 10^{-5}$    &  1.283 \\
    1.5 & 1.6 & $2.81 \times 10^{-5}$  &  1.321\\
    1.6 & 1.7 & $1.05 \times 10^{-5}$   & 1.360 \\
    \hline 
    \end{tabular}
    \caption{Same as \autoref{tab:desiBGS} for the ELG galaxy sample  of DESI.}\label{tab:desiELG}
\end{table}

\begin{table}[htbp]
	\centering
   \begin{tabular}{cccc}
			\multicolumn{4}{c}{SKAO HI-galaxy survey}\\
			\hline
			\hline
			$z_{\rm min}$ & $z_{\rm max}$ & $\bar N_i\,[\mathrm{Mpc}^{-3}]$ & $b_{\rm g}$ \\
			\hline
			0.0 & 0.1 & $2.73 \times 10^{-2}$ & 0.657  \\
			0.1 & 0.2 & $4.93 \times 10^{-3}$ & 0.714  \\
			0.2 & 0.3 & $9.49 \times 10^{-4}$ & 0.789  \\
			0.3 & 0.4 & $2.23 \times 10^{-4}$ & 0.876 \\
			0.4 & 0.5 & $6.44 \times 10^{-5}$ & 0.966 \\
			\hline
		\end{tabular}
	\caption{\label{tab:contredshiftbins} Same as \autoref{tab:desiELG} for the HI galaxy sample of SKAO. 
 }\label{tab:SKAOHI}
\end{table}
\FloatBarrier

\section{IM specifications}\label{sec:IMspecs}
\begin{table}[htbp]
	\centering
	\begin{tabular}{ccccc}
	\multicolumn{5}{c}{SKAO IM survey}\\
		\hline
		\hline
		$D_{\rm d}\,[\mathrm{m}]$ & $f_{\rm sky}$ & $t_{\rm tot}\,[\mathrm{hr}]$ & $N_{\rm d}$ & $T_{\rm sys}$ \\
		\hline
		$15$ & $0.48$ & $10\,000$ & $197$ & Table~7 of \citep{2020arXiv200906197J} \\
		\hline
	\end{tabular}
		\caption{SKAO IM survey specifications used for computing the instrumental noise and the beam effect. We show here the dish diameter $D_{\rm d}$, the sky fraction $f_{\rm sky}$, the observation time $t_{\rm tot}$, the number of dishes $N_{\rm d}$, and the system temperature $T_{\rm sys}$.} \label{tab:IMspecsNoise} 
\end{table}

\FloatBarrier

\section{Angular probes specifications}\label{sec:angularspecs}
\begin{table}[htbp]
	\centering
	\begin{tabular}{cccccc}
		 \multicolumn{6}{c}{WL surveys} \\
		\hline
		\hline 
         & $f_{\rm sky}$ & $\sum_i\,\bar n_i$ & $\epsilon_{\rm int}$ & $N_{\rm b}$ & $\sigma_z^{\rm ph}/(1+z)$ \\
          & & $[\mathrm{arcmin}^{-2}]$ & & & \\
		\hline
         SKAO & $0.1166$ & $3.1$ & $0.22$  & $10$ & $0.05$ \\
         VRO & $0.35$ & $27$ & $0.26$ & $10$ & $0.05$ \\
		\hline 
	\end{tabular}
	\caption{ \label{tab:ska-wl1} Specifications for the WL surveys with SKAO \citep{harrison_ska_2016} and VRO \citep{LSST:2008ijt}.
  }
\end{table}

\bibliographystyle{elsarticle-num-names}
\bibliography{1-MG-CAMB-Cosmo}

\end{document}